\documentclass[a4paper,11pt,pdftex]{article}
\pdfoutput=1 

\usepackage{jheppub} 

\usepackage[utf8]{inputenc}
\usepackage{amsfonts}
\usepackage{amsbsy}
\usepackage{url}
\usepackage{enumerate}
\usepackage[justification=raggedright]{caption}
\usepackage{hhline}

\usepackage{graphicx,color}
\usepackage{amssymb,amsmath}
\usepackage{slashed}
\usepackage{hyperref}
\usepackage{braket}
\usepackage{simplewick}
\usepackage{ascmac}
\usepackage{latexsym}
\usepackage{pifont}          
\usepackage{bm}
\usepackage{comment}
\usepackage[normalem]{ulem}

\newcommand{\n}{\nonumber}

\newcommand{\mr}[1]{\mathrm{#1}}

\newcommand{\h}[1]{\hspace{#1}}

\newcommand{\f}[2]{\frac{#1}{#2}}

\makeatletter
\gdef\@fpheader{}
\makeatother

\makeatletter
\@addtoreset{equation}{section}
\makeatother

\begin{document}
\preprint{YGHP-21-4, KEK-TH-2370}

\title{Stable $Z$-strings with topological polarization in two Higgs doublet model}

\author[a,b]{Minoru~Eto,}
\author[c]{Yu~Hamada,}
\author[b,d]{and Muneto~Nitta} 

\affiliation[a]{Department of Physics, Yamagata University, Kojirakawa-machi 1-4-12, Yamagata, Yamagata 990-8560, Japan}
\affiliation[b]{Research and Education Center for Natural Sciences, Keio University, 4-1-1 Hiyoshi, Yokohama, Kanagawa 223-8521, Japan}
\affiliation[c]{KEK Theory Center, Ibaraki, Japan}
\affiliation[d]{Department of Physics, Keio University, 4-1-1 Hiyoshi, Kanagawa 223-8521, Japan}

\emailAdd{meto(at)sci.kj.yamagata-u.ac.jp}
\emailAdd{yuhamada(at)post.kek.jp}
\emailAdd{nitta(at)phys-h.keio.ac.jp}

\abstract{
We find that a $Z$-string is stable in a wide range of parameter space of the two Higgs doublet model due to 
a split into a pair of two topological $Z$-strings with fractional $Z$ fluxes.
This configuration, a bound state of the two strings connected by a domain wall, is called a vortex molecule.
Although the vortex molecule has no net topological charge,
the locally induced topological charge density is polarized, namely distributed positively around one constituent string 
and negatively around the other constituent string,
leading to the stability of the molecule.
We numerically show that the vortex molecule is indeed a stable solution of the equation of motions
in a much wider parameter space of the model than 
the usual axially symmetric $Z$-string in the Standard Model and the two Higgs doublet model,
although it is not the case for experimental values of the parameters.
}

\maketitle

\section{Introduction}

Vortex strings (or cosmic strings) \cite{Vilenkin:2000jqa,Hindmarsh:1994re,Manton:2004tk,Vachaspati:2015cma} are topological defects 
depending on two spatial directions formed by spontaneous symmetry breaking 
and have been attracted much attention in various contexts of physics
 from elementary particle physics, quantum field theories, nuclear physics, 
 cosmology and astrophysics to condensed matter physics.
For instance, in cosmology, they were expected to contribute to 
the anisotropy of the cosmic microwave background \cite{Vilenkin:2000jqa,Ringeval:2010ca,Hindmarsh:2018wkp} 
and galaxy structure formation \cite{Vilenkin:2000jqa,Shlaer:2012rj,Duplessis:2013dsa}.
They also play important roles in various condensed matter systems \cite{Volovik,Pismen},
supersymmetric field theories \cite{Auzzi:2003fs,Auzzi:2003em,Hanany:2003hp,Hanany:2004ea,Shifman:2004dr,Eto:2005yh,Eto:2006cx,Tong:2005un,Eto:2006pg,Shifman:2007ce,Shifman:2009zz}
and dense QCD matter \cite{Eto:2013hoa}.
One of the simplest field theoretical models of vortex strings is the Abrikosov-Nielsen-Olesen (ANO) vortex \cite{Abrikosov:1956sx,Nielsen:1973cs}
in the Abelian-Higgs model relevant to conventional superconductors,
in which a spontaneously broken $U(1)$ gauge symmetry yields the vacuum a non-trivial first homotopy group $\pi_1 [U(1)] \simeq \mathbb{Z}$, 
leading to the topological stability of the ANO vortex.

Vortex strings are also discussed in the Standard Model (SM) of elementary particles.
The ANO vortex can be embedded into the electroweak (EW) sector of SM  
and are called the $Z$-string \cite{Nambu:1977ag,Vachaspati:1992fi} (see Ref.~\cite{Achucarro:1999it} for a review),
which contains a quantized magnetic flux tube of the $Z$ gauge field.
Many people have studied cosmological consequences of the $Z$-strings;
for instance, EW baryogenesis via the $Z$-strings \cite{Brandenberger:1992ys,Barriola:1994ez,Vachaspati:1994ng}
and production of primordial magnetic fields by the Nambu monopoles \cite{Nambu:1977ag} terminating a $Z$-string \cite{Vachaspati:2001nb,Poltis:2010yu}.
However, because of the lack of non-trivial topology in the SM (the vacuum manifold 
$[SU(2)_W \times U(1)_Y]/U(1)_{\rm EM} \simeq S^3$ has a trivial first homotopy group $\pi_1(S^3)\simeq 0$), 
the $Z$-string can be stable only when the Weinberg angle $\theta_W$ is close to $\pi/2$ \cite{James:1992zp,James:1992wb,Goodband:1995he},
and hence is unstable for the experimental value of the Weinberg angle $\sin^2 \theta_W \simeq 0.23$.
To stabilize it, various mechanisms have been proposed thus far,
for instance, thermal effect \cite{Holman:1992rv,Nagasawa:2002at}, 
fermions trapped around the $Z$-strings \cite{Moreno:1994bk,Earnshaw:1994jj,Naculich:1995cb,Starkman:2000bq,Starkman:2001tc}, and
external magnetic fields \cite{Garriga:1995fv}
(see also Refs.~\cite{Chatterjee:2018znk,Forgacs:2019tbn}).

The two Higgs doublet model (2HDM) is one of the popular and natural extensions of the SM 
(see, \textit{e.g.}, Refs.~\cite{Gunion:1989we,Branco:2011iw} for a review)
and is well motivated for the realization of the EW baryogenesis \cite{Kuzmin:1985mm}
and supersymmetric extensions of the SM \cite{Haber:1984rc}.
It introduces two Higgs doublet fields, instead of one in the case of the SM. 
In spite of its simpleness, phenomenology of the model is quite rich 
thanks to the existence of four additional scalar degree of freedom 
(charged Higgs bosons ($H^{\pm}$), CP-even Higgs boson ($H^0$) and CP-odd Higgs boson ($A^0$)), 
which can be in principle produced at the LHC
(see, e.g., Refs.~\cite{Kanemura:2014bqa,Kanemura:2015mxa,Haller:2018nnx,Bernon:2015qea} and references therein).
As in the case of the SM, 
one can consider an embedding of the ANO vortex into the EW sector of the 2HDM 
\cite{Earnshaw:1993yu,Perivolaropoulos:1993gg} (see also Ref.~\cite{Ivanov:2007de}).
This configuration described by the two Higgs doublet fields with the same winding phases 
is a non-topological solution since it does not have any topological charge 
as the $Z$-string in the SM.
We call this string a \textit{non-topological $Z$-string} in the 2HDM.
The non-topological string is unstable for $\sin ^2 \theta_W \simeq 0.23$ as well as the $Z$-string in the SM. 
On the other hand, the 2HDM can admit a nontrivial topology of the vacuum \cite{Battye:2011jj,Brawn:2011}
by imposing a global $U(1)_a$ symmetry on the Higgs potential, which is spontaneously broken by the Higgs VEVs.
This allows the existence of a topologically stable vortex string containing a fractionally quantized $Z$ flux \cite{La:1993je,Dvali:1993sg,Dvali:1994qf,Battye:2011jj,Eto:2018hhg,Eto:2018tnk} 
(see also Ref.~\cite{Bimonte:1994qh} for a $U(1)_a$-gauged model),
which we call \textit{the topological $Z$-string} in the 2HDM. 
While the topological $Z$-string is topologically stable,
the $U(1)_a$ symmetry should be explicitly broken to give masses to additional Higgs fields, 
attaching domain walls to the topological $Z$ string \cite{Eto:2018hhg,Eto:2018tnk}
as axion strings. 
There are also other studies in the literature on (non-)topological solitons in the 2HDM;
domain walls and sine-Gordon solitons \cite{Bachas:1995ip,Riotto:1997dk,Battye:2011jj,Battye:2020sxy,Battye:2020jeu,Chen:2020soj,Law:2021ing},
sphaleron(-like) solution \cite{Bachas:1996ap,Grant:1998ci,Grant:2001at,Brihaye:2004tz},
global monopoles \cite{Battye:2011jj}
and the Nambu monopoles \cite{Eto:2019hhf,Eto:2020hjb,Eto:2020opf}.

In this paper, 
we show that the non-topological $Z$-string in the 2HDM can be stable 
when it is polarized into a pair of two topological $Z$-strings with fractionally quantized $Z$-fluxes.
This polarized configuration is not axially symmetric and regarded as a bound state of the two topological strings connected by a domain wall.
We call it a \textit{vortex molecule}.
Since it belongs to the same topological sector as a single non-topological $Z$-string, 
the total topological charge of this molecule is zero.
Nevertheless, a topological charge density is locally induced 
around the two constituent strings, and thus we call it ``topological polarization''.
First, we find a sufficient condition on the model parameters in an analytic way
for the vortex molecule to be a stable solution of the equations of motion (EOMs).
When this condition is satisfied, the molecule does not shrink into a single non-topological $Z$-string
due to a repulsive force between the two constituent strings.
Furthermore, the distributions of the topological charge that are locally induced around the two constituent strings
do not have overlap with each other, which means that the topological charge is significantly polarized,
resulting in that the each constituent string does not decay individually 
due to the conservation of the topological charge.
This leads to the stability of the vortex molecule.
Second, we perform numerical simulations confirming the existence of the stable molecule by
using the relaxation method, and determine the parameter region that the non-topological $Z$ string is stable.
The molecule is stable \textit{even without} the $U(1)_a$ symmetry in the Higgs potential,
resulting in a domain wall stretching between the two constituent strings.
Therefore the vortex molecule is stable for a much wider parameter range
compared to the usual axially symmetric non-topological $Z$ strings studied in Refs.~\cite{James:1992zp,Goodband:1995he,James:1992wb,Holman:1992rv,Earnshaw:1993yu,Garriga:1995fv,Nagasawa:2002at,Forgacs:2019tbn}.

In the SM, 
$Z$ strings reduce to semilocal strings in the limit of $\theta_W = \pi/2$ in which 
the $SU(2)_W$ weak gauge symmetry becomes a global symmetry 
\cite{Vachaspati:1991dz,Achucarro:1999it}. 
Semilocal strings are also nontopological strings, and in this case 
a stabilization mechanism similar to ours was proposed 
by adding an $SU(2)$ symmetry breaking term,   
where a single semilocal string is split into two fractional strings 
\cite{Eto:2016mqc}. 
However, in this case, the breaking term introduces an additional topology to the system 
(the vacuum manifold reduces from $S^3$ to $S^1\times S^1$)
turning the semilocal strings to topological strings, 
in which case it is different from our mechanism.

This paper is organized as follows.
In Sec.~\ref{sec:model}, the setup and our notation are presented.
Some definitions for symmetries and basis of the doublets are also given.
In Sec.~\ref{sec:EWstrings}, we give a brief review on the topological and non-topological $Z$-strings in the 2HDM.
In Sec.~\ref{sec:interaction} we study interaction between the two topological $Z$-strings.
We illustrate that there are four attractive/repulsive forces.
In Sec.~\ref{sec:analytic-molecule}, we discuss the stability of the vortex molecule in an analytic way.
A sufficient condition of the stability is obtained.
In Sec.~\ref{sec:numerical-molecule}, we solve the EOMs numerically 
and confirm the existence of the stable vortex molecule.
We show the parameter space for the stability of the molecule.
In Sec.~\ref{sec:conclusion}, we give a summary and discussion.
In Appendix \ref{app:basis}, difference between the notations of this paper and our previous paper
is summarized for readers' convenience.
In Appendix \ref{app:stability-Zstring}, a review on the (in)stability of the $Z$-string in the SM is presented.
In Appendix \ref{app:interaction}, we give detailed computations of the interaction of the topological $Z$-strings.

\section{The 2HDM}\label{sec:model}
\subsection{The Lagrangian and Higgs potential}
We introduce two $SU(2)$ doublets of Higgs scalar fields, $\Phi_1$ and $\Phi_2$, both with the hypercharge $Y=1$.
The Lagrangian which describes the gauge and Higgs sectors can be written as
\begin{align}
\h{-1em} {\mathcal L} = - \frac{1}{4}\left(Y_{\mu\nu}\right)^2 - \frac{1}{4}\left(W_{\mu\nu}^a\right)^2  + \left|D_\mu \Phi_i \right|^2 - V(\Phi_1, \Phi_2).
\label{eq:L}
\end{align}
Here, $Y_{\mu\nu}$ and $W^a_{\mu\nu}$ describe field strength tensors of the hypercharge
and weak gauge interactions, respectively, with $\mu$ ($\nu$) and $a$ being Lorentz and weak iso-spin indices, respectively. 
$D_\mu$ represents the covariant derivative acting on the Higgs fields, and the index $i$ runs $i=1,2$.
The most generic quartic potential $V(\Phi_1, \Phi_2)$ for the two Higgs doublets is given by
\begin{align}
V(\Phi_1,\Phi_2) & = m_{11}^2\Phi_1^\dagger\Phi_1 + m_{22}^2 \Phi_2^\dagger \Phi_2 
 - \left(m_{12}^2 \Phi_1^\dagger \Phi_2 + {\rm h.c.}\right) \n\label{eq:potential} \\
 & + \frac{\beta_1}{2}\left(\Phi_1^\dagger\Phi_1\right)^2 + \frac{\beta_2}{2}\left(\Phi_2^\dagger\Phi_2\right)^2 \nonumber\\
& + \beta_3\left(\Phi_1^\dagger\Phi_1\right)\left(\Phi_2^\dagger\Phi_2\right) 
   + \beta_4 \left(\Phi_1^\dagger\Phi_2\right)\left(\Phi_2^\dagger\Phi_1\right)\n  \\
 &+ \left\{ \left(
\frac{\beta_5}{2}\Phi_1^\dagger\Phi_2
+\beta_6 |\Phi_1|^2
+\beta_7 |\Phi_2|^2
\right)\Phi_1^\dagger\Phi_2
+ {\rm h.c.}\right\} \, .
\end{align}
In this paper, we assume that both the Higgs fields develop real vacuum expectation values (VEVs) as \footnote{
Note that this convention is different by the factor ``$\sqrt{2}$'' from the usual notation in the literature.}
\begin{equation}
 \Phi_1 = 
\begin{pmatrix}
 0 \\ v_1
\end{pmatrix},\quad 
\Phi_2 =  
\begin{pmatrix}
 0 \\ v_2
\end{pmatrix}.
\end{equation}
Then the electroweak scale, $v_\mr{EW}$ ($\simeq $ 246 GeV), can be expressed by these VEVs as $v_{\rm EW}^2 = 2 v_{\rm sum}^2 \equiv 2 (v_1^2 + v_2^2)$.
The masses of the gauge bosons are given by
\begin{equation}
 m_W^2 = \frac{g^2v_\mr{EW}^2}{4}, \quad  m_Z^2 = \frac{g^2 v_\mr{EW}^2}{4 \cos ^2 \theta_W}
\end{equation}
with the standard definitions $\cos \theta_W \equiv \frac{g}{\sqrt{g^2 + g'^2}}$,
$Z_\mu \equiv W_\mu^3 \cos \theta_W -Y_\mu \sin \theta_W$ and
$A_\mu \equiv W_\mu^3 \sin \theta_W +Y_\mu \cos \theta_W$.
The $W^\pm$ bosons are defined as $ W_\mu^\pm \equiv (W_\mu^1 \pm i W_\mu^2)/\sqrt{2}$.

For later use, we rewrite the Higgs fields in a two-by-two matrix form \cite{Grzadkowski:2010dj}, $H$,
defined by 
\begin{equation}
 H = \left( i\sigma_2 \Phi_1^*,\ \Phi_2\right).
\end{equation}
The matrix scalar field $H$ transforms under the electroweak $SU(2)_W \times U(1)_Y$ symmetry as
\begin{equation}
H \to \exp\left[\f{i}{2}\theta_a(x) \sigma_a\right] H ~\exp\left[-\f{i}{2} \theta_Y(x) \sigma_3\right],
\end{equation}
where the group element acting from the left belongs to $SU(2)_W$ and the other element acting from the right beongs to $U(1)_Y$.
Therefore the covariant derivative on $H$ can be expressed as
\begin{equation}
D_\mu H =\partial_\mu H - i \frac{g}{2} \sigma_a W_\mu^a H + i \frac{g'}{2}H\sigma_3 Y_\mu.
\end{equation}
The VEV of $H$ is expressed by a diagonal matrix $\langle H \rangle = \mr{diag} (v_1,v_2)$,
and the Higgs potential can be rewritten by using $H$ as follows:
\begin{align}
V(H)
& = - m_{1}^2~ \mr{Tr}|H|^2 - m_{2}^2~ \mr{Tr}\left(|H|^2 \sigma_3\right) - \left( m_{3}^2 \det H + \mr{h.c.}\right)\nonumber\label{eq:potential2} \\
& + \alpha_1~\mr{Tr}|H|^4  +  \alpha_2 ~\left(\mr{Tr}|H|^2 \right)^2+ \alpha_3~ \mr{ Tr}\left(|H|^2 \sigma_3 |H|^2\sigma_3\right)  \n \\
& + \alpha_4~ \mr{Tr}\left(|H|^2 \sigma_3 |H|^2\right) \n \\
& + \left\{\left(
\alpha_5 \det H + \alpha_6 \mr{Tr}|H|^2 + \alpha_7 \mr{Tr}\left(|H|^2 \sigma_3 \right) 
\right) \det H +\mr{h.c.}
\right\},
\end{align}
with $|H|^2 \equiv H ^\dagger H$.
The parameters $m_3$, $\alpha_5$, $\alpha_6$, and $\alpha_7$ are complex in general.
The parameter sets in the Higgs potential in Eqs.~\eqref{eq:potential} and \eqref{eq:potential2}
are related as
\begin{align}
  m_{11}^2 = -m_1^2 - m_2^2 , \h{2em}  m_{22}^2 = -m_1^2 + m_2^2 ,  \h{2em}  m_{12}^2 = m_3^2 ,
  \nonumber \\
  \beta_1  =2(\alpha_1+ \alpha_2 + \alpha_3 +  \alpha_4 ), \h{2em} \beta_2  =2( \alpha_1+ \alpha_2 + \alpha_3 -  \alpha_4 ), \nonumber \\
  \beta_3 = 2(\alpha_1+ \alpha_2 - \alpha_3) ,  \h{2em} \beta_4 = 2(\alpha_3 - \alpha_1),  \h{2em} \beta_5 = 2\alpha_5, \nonumber \\
\beta_6 = \alpha_6 + \alpha_7, \h{2em} \beta_7 = \alpha_6 -\alpha_7 \, .
\end{align}

\subsection{Custodial symmetry and CP symmetry}
By the electroweak precision measurement,
the electroweak $\rho$ parameter,
defined by 
$\rho\equiv m_W^2 / m_Z^2 \cos^2 \theta_W$,
must be close to unity \cite{ParticleDataGroup:2020ssz}.
To satisfy this experimental constraint,
we impose a global $SU(2)_L\times SU(2)_R$ symmetry on the Higgs potential.
It is convenient to introduce the following two matrices:
\begin{equation}
 H_1 \equiv \left( i\sigma_2 \Phi_1^*,\ \Phi_1\right), \quad  H_2 \equiv \left( i\sigma_2 \Phi_2^*,\ \Phi_2\right) \, ,
\end{equation}
which transform under $SU(2)_L\times SU(2)_R$ as
\begin{equation}
  H_1 \to L H_1 R^\dagger, \quad  H_2 \to L H_2 R^\dagger, \hspace{2em} L,R \in SU(2)_{L,R} \label{eq:SU2L_SU2R} \,.
\end{equation}
The potential $V(\Phi_1,\Phi_2)$ given in Eq.~\eqref{eq:potential} is invariant under this transformation when 
\begin{equation}
\beta_4 = \beta_5, \quad m_{12}^2, \beta_5 , \beta_6, \beta_7 \in \mathbb{R} \, ,
\end{equation}
which, in the notation in Eq.~\eqref{eq:potential2}, is equivalent to 
\begin{equation}
\alpha_3 - \alpha_1 = \alpha_5,  \quad m_3, \alpha_5 , \alpha_6, \alpha_7 \in \mathbb{R} \,.\label{eq:cond_custodial1}
\end{equation}
By the Higgs VEVs, 
$SU(2)_L\times SU(2)_R$ symmetry is spontaneously broken,
but its diagonal subgroup
\begin{equation}
  H_1 \to U H_1 U^\dagger, \quad  H_2 \to U H_2 U^\dagger, \hspace{2em} U \in SU(2)_\mr{C} \label{eq:custodial1} \,
\end{equation}
remains if $v_1, v_2 \in \mathbb{R}$.
This residual symmetry is called the custodial $SU(2)_\mr{C}$ symmetry.
This symmetry forces the CP-odd Higgs boson and the charged Higgs boson to be degenerate,
so that the $\rho$ parameter is protected from receiving their large loop corrections.
Note that the kinetic term of $H$ cannot be invariant under this transformation
because of the presence of the $U(1)_Y$ gauge field.
Thus the custodial symmetry is not an exact symmetry of the theory 
but is explicitly broken by the $U(1)_Y$ gauge interaction.

The bosonic Lagrangian ${\mathcal L}$ in Eq.~(\ref{eq:L}) 
is invariant under a $\mathbb{Z}_2$ transformation defined by 
\begin{align} 
 \begin{cases}
  H_i \to i\sigma_2 H_i (i \sigma_2)^\dagger  \\
  W_\mu \to  i\sigma_2 W_\mu (i \sigma_2)^\dagger \\
  Y_\mu \to - Y_\mu \, , 
 \end{cases} \label{eq:CP1}
\end{align}
for $i=1,2$
when the parameters of the potential are real:
\begin{equation}
\text{CP conserving condition}:\quad 
m_3, \alpha_5 , \alpha_6, \alpha_7 \in \mathbb{R}
\label{eq:cond_real} \, .
\end{equation}
This is nothing but the CP symmetry (more precisely, C symmetry for the gauge fields)
and acts on $H_i$ as a subgroup of the custodial $SU(2)_\mr{C}$ transformation since $i \sigma_2 \in SU(2)_\mr{C}$. 
(Correspondingly, Eq.~\eqref{eq:cond_real} is a necessary condition of Eq.~\eqref{eq:cond_custodial1}.)
Thus the custodial $SU(2)_\mr{C}$ symmetric Higgs potential is automatically invariant under the  CP symmetry.
In the doublet notation, the CP transformation acting on the Higgs fields is expressed as the complex conjugation:
\begin{align}
 \begin{cases}
  \Phi_1 \to \Phi_1 ^\ast \\
  \Phi_2 \to \Phi_2 ^\ast  \, ,
\end{cases} \label{eq:CP-doublet-1}
\end{align}
and the CP symmetry is preserved in the vacuum if $v_1, v_2 \in \mathbb{R}$.

\subsection{Basis transformation}
For later use, we move to another basis of the Higgs fields.
It is known that
2HDM Lagrangian without the Yukawa couplings has an ambiguity corresponding to the $U(2)$ basis transformation of the Higgs doublets 
(see, \textit{e.g.}, Refs.~\cite{Davidson:2005cw,Haber:2006ue,Grzadkowski:2010dj,Haber:2010bw} and Ref.~\cite{Branco:2011iw} for a review):
\begin{equation}
\Phi_i \to \Phi_i' = \sum_{j=1}^2M_{ij} \Phi_j, \quad M \in U(2) \quad (i=1,2) \, .\label{eq:basis}
\end{equation}
All Higgs potentials that are related by this basis transformation are physically equivalent and predict the same physics.
Let us take the matrix $M_{ij}$ as
\begin{equation}
 M= \frac{1}{2}
\begin{pmatrix}
 1-i & 1+i \\
 1+i & 1-i
\end{pmatrix} \, .\label{eq:basis-trsf}
\end{equation}
By this basis transformation in Eq.~\eqref{eq:basis}, 
the Higgs potential can be rewritten in terms of the new doublets $\Phi_1'$ and $\Phi_2'$ as
\begin{align}
V(\Phi_1,\Phi_2) 
= & - m_{1}'^2~ \mr{Tr}|H'|^2 - m_{2}'^2~ \mr{Tr}\left(|H'|^2 \sigma_3\right) - \left( m_{3}'^2 \det H' + \mr{h.c.}\right)\nonumber \\
& + \alpha_1'~\mr{Tr}|H'|^4  +  \alpha_2' ~\left(\mr{Tr}|H'|^2 \right)^2+ \alpha_3' ~ \mr{ Tr}\left(|H'|^2 \sigma_3 |H'|^2\sigma_3\right)  \n \\
& + \alpha_4'~ \mr{Tr}\left(|H'|^2 \sigma_3 |H'|^2\right) \n \\
& + \left\{\left(
\alpha_5' \det H' + \alpha_6' \mr{Tr}|H'|^2 + \alpha_7' \mr{Tr}\left(|H'|^2 \sigma_3 \right) 
\right) \det H' +\mr{h.c.}
\right\},\label{eq:potential_new_basis}
\end{align}
with $ H' = \left( i\sigma_2 \Phi_1'^*,\ \Phi_2'\right)$.
Here we have put the prime $'$ on the parameters to distinguish them from ones in the old basis in Eq.~\eqref{eq:potential2}.
The kinetic term of $H$ is simply $\mathrm{Tr}|D_\mu H'|^2$
since the new doublets have the same gauge charges as those of the old ones.
One should note that the parameters $m_3', \alpha_5', \alpha_6', \alpha_7'$ are not real in the new basis
even when the CP symmetry is imposed in the old basis in Eq.~\eqref{eq:cond_real}.
The explicit relation between the parameters in the old Higgs potential in Eq.~\eqref{eq:potential} and the new one in Eq.~\eqref{eq:potential_new_basis} is given as
\begin{align}
 m_{11}^2 &=  - m_1'^2 + \mr{Im}\, m_3'^2  \, ,~ \n \\
 m_{22}^2 &=  - m_1'^2 - \mr{Im}\, m_3'^2 \, ,\n \\
 m_{12}^2 & =  i m_2'^2 + \mr{Re}\, m_3'^2  \, , \n \\
\beta_1 & = \alpha_1' + 2\alpha_2' + \alpha_3' - \mr{Re}\, \alpha_5' - 2 \, \mr{Im}\, \alpha_6' \, ,\n \\
\beta_2 & = \alpha_1' + 2\alpha_2' + \alpha_3' - \mr{Re}\, \alpha_5' + 2 \, \mr{Im}\, \alpha_6' \, , \n \\
\beta_3 & = 3\alpha_1' + 2 \alpha_2' - \alpha_3' + \mr{Re}\, \alpha_5' \, , \n \\
\beta_4 & = -\alpha_1' +  3\alpha_3' + \mr{Re}\, \alpha_5' \, , \n \\
\beta_5 & = -\alpha_1' - \alpha_3' + \mr{Re}\, \alpha_5' +2i \,\mr{Re}\, \alpha_7' \, , \n \\
\beta_6 & = i\alpha_4' - \mr{Im} \, \alpha_5' + \mr{Re}\, \alpha_6' - i \,\mr{Im}\, \alpha_7' \, , \n \\
\beta_7 & = i\alpha_4' + \mr{Im} \, \alpha_5' + \mr{Re}\, \alpha_6' + i \,\mr{Im}\, \alpha_7' \, .
\end{align}

In this new basis, the $SU(2)_L\times SU(2)_R$ symmetry can be expressed as \cite{Grzadkowski:2010dj,Pomarol:1993mu}
\begin{equation}
  H' \to L H' R^\dagger, \quad  L,R \in SU(2)_{L,R} \label{eq:SU2L_SU2R_2} \,.
\end{equation}
The Higgs potential $V(H')$ (Eq.~\eqref{eq:potential_new_basis}) is invariant under this $SU(2)_L\times SU(2)_R$ symmetry when 
\begin{equation}
SU(2)_L\times SU(2)_R: \quad m_2'=\alpha_3'=\alpha_4'=\alpha_7'=0 .\label{eq:cond_2}
\end{equation}
Note that this condition of the $SU(2)_L\times SU(2)_R$ symmetry seems different from Eq~\eqref{eq:cond_custodial1}
although they are physically equivalent \cite{Davidson:2005cw,Haber:2006ue,Grzadkowski:2010dj,Haber:2010bw}.
This symmetry is spontaneously broken into its subgroup $SU(2)_\mr{C}$,
\begin{equation}
H' \to U' H' U'^\dagger, \hspace{2em} U' \in SU(2)_\mr{C} \,  \label{eq:custodial2}
\end{equation}
if $v_1'^\ast = v_2'$.
This is the custodial $SU(2)_\mr{C}$ symmetry in the new basis
and equivalent to that in the old basis.
This corresponds to the case II studied by Pomarol and Vega in Ref.~\cite{Pomarol:1993mu}.
The CP symmetry, which acts on $H'$ as a $\mathbb{Z}_2$ subgroup of $SU(2)_\mr{C}$,
can be expressed in this basis as
\begin{align}
 \begin{cases}
  H' \to i\sigma_2 H' (i \sigma_2)^\dagger \\
  W_\mu \to  i\sigma_2 W_\mu (i \sigma_2)^\dagger \\
  Y_\mu \to - Y_\mu \, ,
 \end{cases} \label{eq:CP2}
\end{align}
or, in terms of the doublet notation for the Higgs field, 
\begin{align}
 \begin{cases}
  \Phi_1' \to \Phi_2'^\ast \\
  \Phi_2' \to \Phi_1'^\ast  \, .
\end{cases} \label{eq:CP-doublet-2} 
\end{align}
Thus, in this basis, the CP transformation is a combination of the complex conjugation and the exchange of the doublets.
Imposing the CP symmetry on the Lagrangian reads
\begin{equation}
\text{CP conserving condition}:\quad m_2' = \alpha_4' = \alpha_7' = 0.
\label{eq:cond_3}
\end{equation}
Note that the CP symmetry is not spontaneously broken in the vacuum if $v_1'^\ast = v_2'$.

Hereafter, we work in this new basis.
Because the old and new basis correspond to the cases I and II studied by Pomarol and Vega \cite{Pomarol:1993mu}, 
we call them ``PV-I basis'' and ``PV-II basis'', respectively.
For a simple notation, we will omit the prime on the parameters and the Higgs fields
even in the PV-II basis.
Note that, in Refs.~\cite{Eto:2019hhf,Eto:2020hjb,Eto:2020opf}, 
Eqs.~\eqref{eq:CP2} and \eqref{eq:CP-doublet-2} (the CP symmetry in the PV-II basis) is called ``$(\mathbb{Z}_2)_\mr{C}$ symmetry''
and the CP symmetry for the Higgs fields is defined in terms of only a charge conjugation instead of Eqs.~\eqref{eq:CP2} and \eqref{eq:CP-doublet-2}.
Correspondingly, some terminology is different from this paper.
See Appendix \ref{app:basis} for summary of their relations.

\subsection{$U(1)_a$ symmetry}
{In Secs.~\ref{sec:EWstrings}, \ref{sec:interaction} and \ref{sec:analytic-molecule}, we impose on the Higgs potential 
in the PV-II basis a global $U(1)$ symmetry,
which is called the $U(1)_a$ symmetry 
and defined by a rotation of the relative phase of the two doublets:
\begin{equation}
 H \to e ^ {i \alpha} H  \quad (0\leq\alpha < 2\pi)\label{eq:U1a}
\end{equation}
or, equivalently,
\begin{equation}
\Phi_1 \to e^{-i\alpha} \Phi_1, \quad \Phi_2 \to e^{i\alpha} \Phi_2 \,.
\end{equation}
The Lagrangian is invariant under the $U(1)_a$ transformation,
when 
\begin{equation}
\text{$U(1)_a$ condition}:\quad m_3 = \alpha_5 = \alpha_6=\alpha_7=0.
\label{eq:cond_1}
\end{equation}
Note that, because the $U(1)_a$ transformation does not commute with the basis transformation in Eq.~\eqref{eq:basis}, 
one should take care about which basis the $U(1)_a$ symmetry is imposed in.
As stated above, we work in the PV-II basis throughout this paper.

After $H$ gets the VEV, this $U(1)_a$ symmetry is spontaneously broken
and the corresponding Nambu-Goldstone (NG) boson appears,
which we call the CP-even Higgs boson ($H^0$).\footnote{This is called CP-odd Higgs boson in Refs.~\cite{Eto:2019hhf,Eto:2020hjb,Eto:2020opf}, see Appendix \ref{app:basis}.}
Note that the CP symmetry is defined by Eq.~\eqref{eq:CP2} in this basis.
As is shown later, the spontaneously broken $U(1)_a$ symmetry gives rise to non-trivial topological excitations.

Because an experimental lower bound on the mass of $H^0$ is typically $\mathcal{O}(100)$ GeV
(which highly depends on how the doublets couple to the SM fermions),
such a massless $H^0$ is phenomenologically disfavored.
Therefore, in realistic models,
we should break the $U(1)_a$ symmetry explicitly by switching on $m_3$, $\alpha_5$, $\alpha_6$ or $\alpha_7$ making $H^0$ massive.

\subsection{Higgs mass spectrum with custodial symmetry}
We present the Higgs mass spectrum 
in the case with the $SU(2)_L \times SU(2)_R$ symmetry,
\textit{i.e.},
\begin{equation}
SU(2)_L\times SU(2)_R: \quad
 m_2= \alpha_3 = \alpha_4 = \alpha_7 = 0 \, 
\end{equation}
while the $U(1)_a$ symmetry is not imposed
because it should be explicitly broken in phenomenologically viable cases.
For simplicity, we suppose that all parameters are real,
\textit{i.e.},
\begin{equation}
 \mr{Im}\, m_3 =  \mr{Im}\, \alpha_5 =  \mr{Im}\, \alpha_6 = 0 \, .
\end{equation}
In this case, the Higgs VEVs become real values given by
\begin{equation}
 v_1 = v_2 = \sqrt{\frac{ m_1^2 + m_3^2}{2(\alpha_1+2\alpha_2+\alpha_5+2\alpha_6)}} \equiv v,\label{eq:vevs}
\end{equation}
($\tan \beta\equiv v_2 / v_1 =1$)
and the custodial symmetry $SU(2)_\mr{C}(\subset SU(2)_L \times SU(2)_R)$ is preserved in the vacuum.

In the matrix notation, fluctuations around the vacuum can be parametrized as
\begin{align}
 H &= v {\bf 1}_2 + \f{1}{\sqrt{2}}\left(\chi^A + i \pi^A \right) \sigma^A  \h{2em} (A=0,\cdots,3)\label{010632_18Mar20}
\end{align}
with $\sigma^A = (\bm{1},\sigma^a)$ ($a=1,2,3$).
Here $\pi^0$ is the CP-even neutral Higgs boson $H^0$.
It becomes the massless Nambu-Goldstone (NG) boson in the case with the exact $U(1)_a$ symmetry under the condition \eqref{eq:cond_1}.
On the other hand, $\pi^a$'s  are would-be NG bosons for $SU(2)_W\times U(1)_Y$ and eaten by the gauge bosons.
In this setup, due to the custodial symmetry, the components $(\chi^1,\chi^2,\chi^3)$ form a custodial triplet
while $\chi^0$ is a custodial singlet.
Taking linear combinations of $\chi^1$ and $\chi^2$, one obtains mass eigenstates with definite $U(1)_\mr{EM}$ charges,
which are identified with the charged Higgs boson $H^\pm$.
On the other hand, $\chi^3$ is neutral under $U(1)_\mr{EM}$ and has odd parity under the CP symmetry defined in Eq.~\eqref{eq:CP2}.
It is called the CP-odd neutral Higgs boson $A^0$.\footnote{In Refs.~\cite{Eto:2019hhf,Eto:2020hjb,Eto:2020opf}, 
this particle is called the CP-even Higgs boson, see Appendix \ref{app:basis}.}
Their masses are degenerate as
\begin{equation}
m_{H^\pm}^2 = m_{A^0}^2 = 2 m_3^2 + 4 v^2 (\alpha_1 - \alpha_5 - \alpha_6) \, .
\end{equation}

Note that $\chi^0$ and $H^0$ do not mix with each other at the tree level and are mass eigenstates
since we set the all parameters as real.
In addition, the CP-even Higgs $H^0$ does not have the $H^0 VV$ coupling with the gauge bosons ($V=W$ and $Z$) at the tree level
while $\chi^0$ has the same coupling with the gauge bosons as that of the SM Higgs boson.
Thus $\chi^0$ must be identified with the SM-like Higgs boson $h^0$.
Their mass eigenvalues are given by 
\begin{equation}
 m_{h^0}^2= 4 v^2 (\alpha_1 + 2 \alpha_2 + \alpha_5 + 2 \alpha_6)
\end{equation}
and
\begin{equation}
 m_{H^0}^2 = 2 m_3^2 - 4 v^2 (2 \alpha_5 + \alpha_6) \, .
\end{equation}

The parameter $m_3^2$ (or $m_{12}^2$) is often called the decoupling parameter
because if $m_3^2$ is taken to infinity $m_3^2 \to \infty$ keeping $v$, 
all particles other than the SM Higgs boson $h^0$ become infinitely heavy,
$m_{H^\pm}^2, m_{H^0}^2, m_{A^0}^2 \to \infty$, 
which means that they are decoupled and that the model reduces to the SM Higgs sector.

\section{Electroweak strings in 2HDM: a review}
\label{sec:EWstrings}
We here give reviews on electroweak strings in the 2HDM.
In this section, unless otherwise noted,
we assume both the $U(1)_a$ and $SU(2)_L\times SU(2)_R$ symmetries,
Eqs.~\eqref{eq:cond_1} and \eqref{eq:cond_2}.
Note that the Higgs VEVs given by Eq.~\eqref{eq:vevs}
spontaneously break the both symmetries, 
the latter of which is broken down to the custodial $SU(2)_\mr{C}(\subset SU(2)_L\times SU(2)_R)$ symmetry.

\subsection{Global $U(1)_a$ string}
Before discussing $Z$-strings, let us point out the simplest string without a $Z$-flux. 
A global $U(1)_a$ string is the simplest string 
whose ansatz is given by
\begin{align}
 H^\text{global} &= v  f^\text{global}( r ) e^{i\varphi} \,  \bm{1}_{2\times 2} ,
 \quad Z_i = 0 \label{eq:H-global},
\end{align}
where $ r \equiv \sqrt{x^2+y^2}$ and $\varphi$ is the azimuthal angle around the $z$-axis. 
The boundary conditions imposed on the profile function are $f^\text{global} (0)=0$ and $f^\text{global}(\infty)=1$.
Thus, the asymptotic form of $H^\text{global}$ at $ r  \to \infty$ 
is $\sim v\, \exp[{i\varphi}] ~\mr{diag} \left(1, 1 \right)$.
This string is accompanied by no $Z$ flux.
Since this is a global vortex, its tension is logarithmically divergent.

In the presence of the $U(1)_a$ symmetry breaking terms, 
the asymptotic potential of the ansatz (\ref{eq:H-global}), 
in which the azimuthal angle $\varphi$ is replaced with a monototicaly increasing function 
$\phi(\varphi)$ of it 
 with the same range ($\phi(0)=0$ and $\phi(2\pi)=2\pi$),
becomes 
\begin{align}
V &\sim (-m_3^2 + 2 \alpha_6 v^2) \det H^\text{global}  
+ \alpha_5 (\det H^\text{global} )^2 + {\rm h.c.} \nonumber\\
&=  2(-m_3^2 + 2 \alpha_6 v^2) v^2 \cos 2\phi + 2\alpha_5 v^2 \cos 4\phi.\label{eq:quad-SG}
\end{align}
This is a variant of sine-Gordon (quadruple sine-Gordon) potential, and thus a single $U(1)_a$ string is attached by 
at most four domain walls \cite{Eto:2018hhg,Eto:2018tnk}. 
How many walls are attached to it depends on the parameters.

\subsection{Non-topological $Z$-string}
For later use, we here review the non-topological $Z$-string obtained by embedding of the $Z$-string in the SM \cite{Nambu:1977ag,Vachaspati:1992fi} into 2HDM.

The non-topological $Z$-string is the same form of the ANO vortex for the two Higgs doublets,
whose ansatz is given as
\begin{align}
 H^\text{non-top} &= v  f^\text{non-top}( r ) e^{i\varphi \sigma_3} \,  \bm{1}_{2\times 2}\label{eq:H-nontop}\\
 Z_i^\text{non-top} &=  - \f{2\cos \theta_\mr{W}}{g}
 \f{\epsilon_{3ij}x^j}{r^2} \left(1- w^\text{non-top}( r )\right) ,\label{eq:Z-nontop}
\end{align}
with the polar coordinates $(r,\varphi)$.
The boundary conditions imposed on the profile functions are $f^\text{non-top} (0)=w ^\text{non-top} (\infty)=0$, $w^\text{non-top} (0)=f^\text{non-top}(\infty)=1$.
Thus, the asymptotic form of $H^\text{non-top}$ at $ r  \to \infty$ 
is $\sim v \exp[{i\varphi}\sigma_3] ~\mr{diag} \left(1, 1 \right)$.
All winding phases of the Higgs field are canceled by the $Z$ gauge field at infinity.
Thus, it is a local vortex whose tension is finite.
The amount of the $Z$ flux is given by
\begin{equation}
 \Phi_Z^\text{non-top} = \frac{4 \pi \cos \theta_\mr{W} }{ g}\label{eq:Zflux-nontop} 
\end{equation}
along the $z$-axis.
This is nothing but that of the non-topological $Z$-string in the SM~\cite{Nambu:1977ag,Vachaspati:1992fi}.

This configuration solves the EOMs as the ANO vortex string.
Note, however, that its stability is not ensured because it does not have any topological charge.
The boundary $S^1$ at $r\to\infty$ is mapped
onto the $U(1)_Z$ gauge orbit, but it can be unwound
inside the whole $SU(2)_W\times U(1)_Y$ gauge orbits.
Indeed, the $Z$-string in the SM is known to be unstable 
for experimental values of the SM Higgs mass, the $W$ boson mass and the Weinberg angle.
See Appendix \ref{app:stability-Zstring}.
The non-topological $Z$-string in the 2HDM is unstable as well due to the same reason.

\begin{figure}[tb]
 \centering
\includegraphics[width=0.45\textwidth]{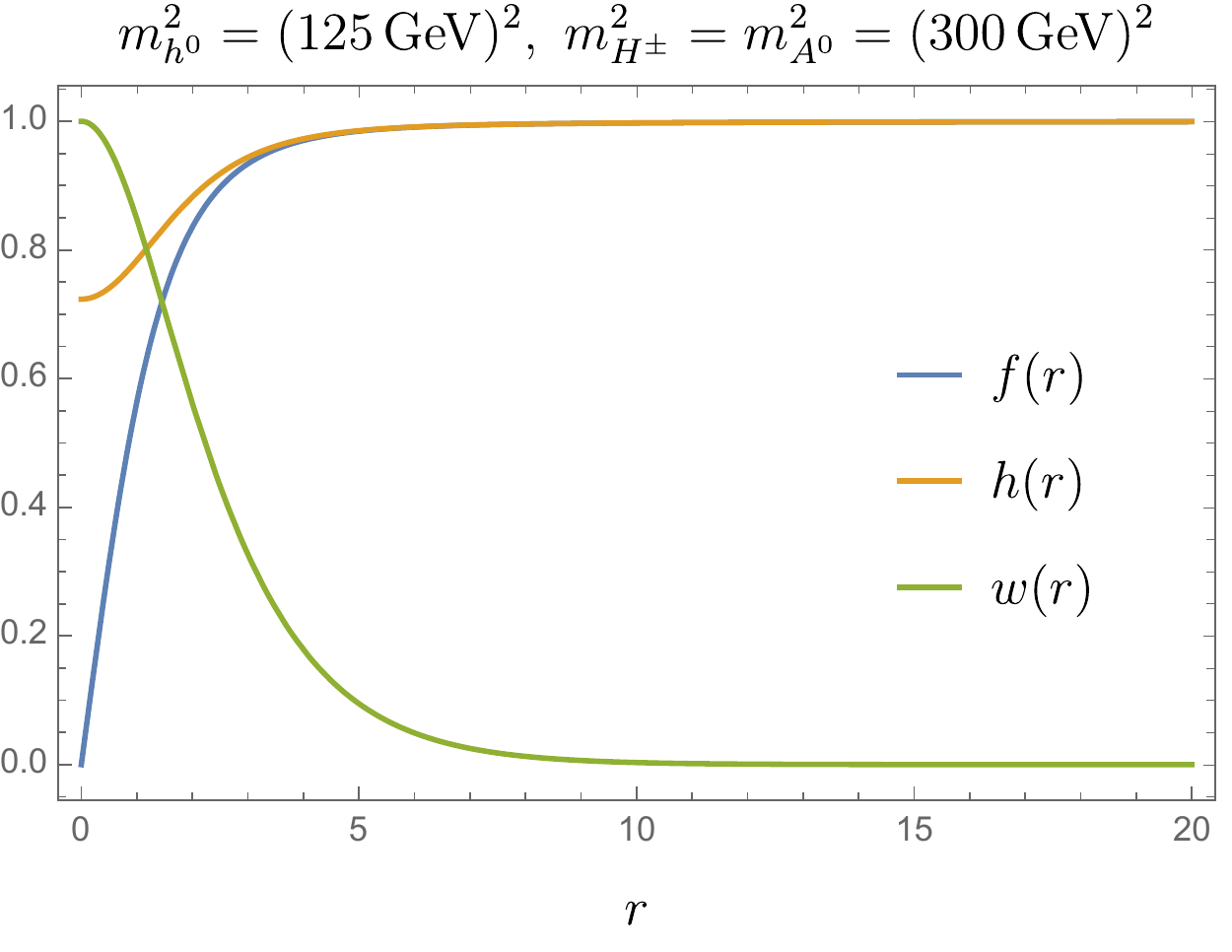} \hspace{2em}
\includegraphics[width=0.45\textwidth]{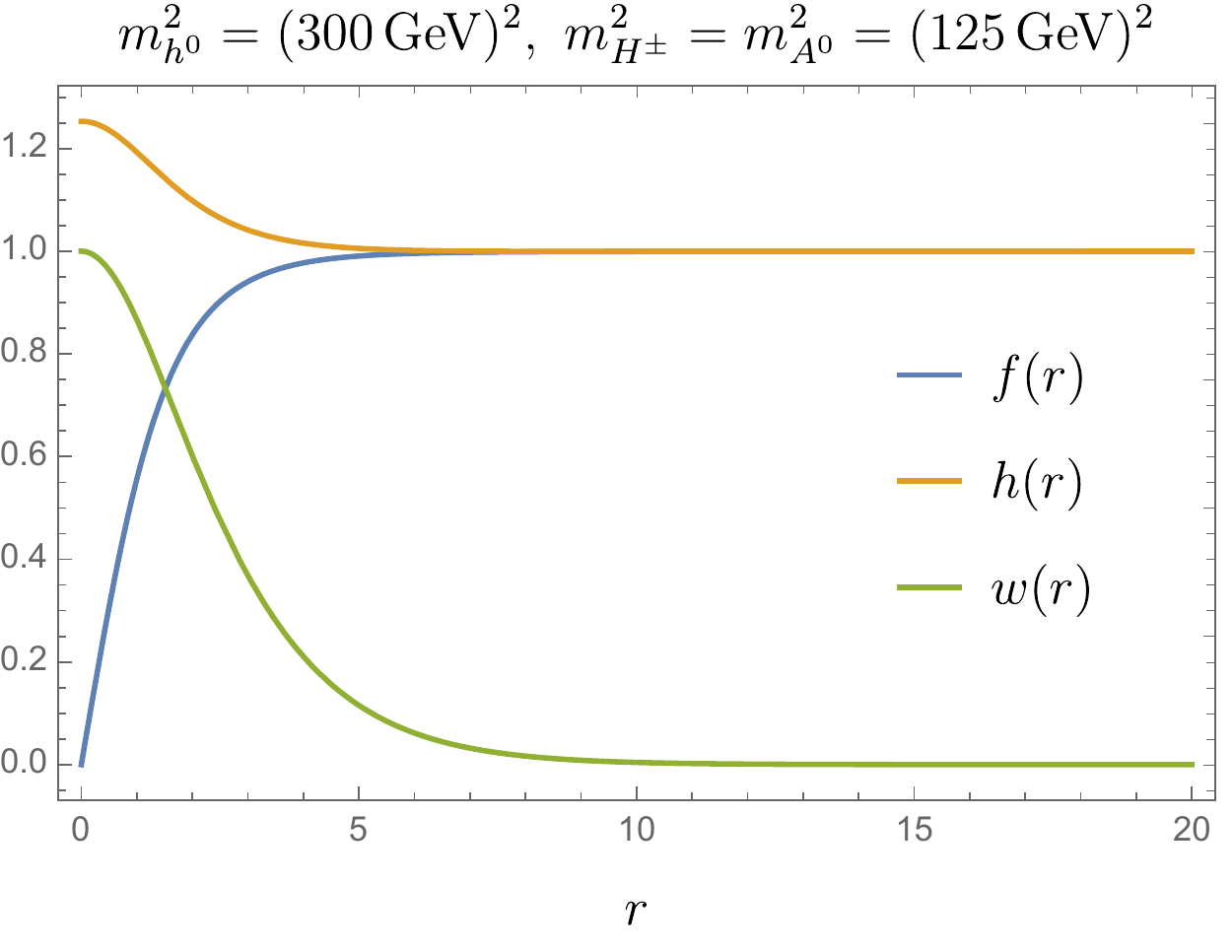} 
\caption{
Plot of the profile functions obtained by solving the EOMs numerically.
The custodial $SU(2)_\mr{C}$ symmetry is imposed.
The masses of the gauge bosons are fixed as $m_Z = 91\, \mathrm{GeV}$ and $m_W = 80\, \mathrm{GeV}$.
We take a length unit such that $v= 123 \, \mathrm{GeV} = 1$.
The condensation of $h(r)$ at the center of the string becomes larger 
when $m_{h^0}$ is larger than the other masses (right panel).
}
\label{fig:10string}
\end{figure}

\subsection{Topological $Z$-strings}
\label{sec:top-Zstring}
In Refs.\cite{Eto:2018tnk,Eto:2018hhg,Dvali:1994qf,Dvali:1993sg}, 
it is pointed out that, unlike in the SM, 
2HDMs allow topologically stable strings to exist thanks to the global $U (1)_a$ symmetry.
First, consider topological strings with the $Z$ flux (topological $Z$-strings).
There are two types of topological $Z$-strings corresponding to which one of the two Higgs doublets has a winding phase.
To see that, let us take $W_\mu^\pm = A_\mu=0$.

One of the topological $Z$-strings is called the $(1,0)$-string,
whose ansatz located on the $z$ axis is given by
\begin{align}
 H^{(1,0)} &= v \begin{pmatrix}
	        f^{(1,0)}( r ) e^{i\varphi} &0 \\
               0 & h ^{(1,0)} ( r )   
	      \end{pmatrix},\label{eq:H(1,0)}\\
 Z_i ^{(1,0)} &=  -  \f{ \cos \theta_\mr{W}}{g}
 \f{ \epsilon_{3ij}x^j}{ r ^2} \left(1- w ^{(1,0)}( r )\right)\label{013304_3Dec19}
\end{align}
with the polar coordinates $(r,\varphi)$.
The boundary conditions imposed on the profile functions are 
\begin{equation}
f ^{(1,0)} (0)={h^{(1,0)}}'(0)=w ^{(1,0)} (\infty)=0, \quad w^{(1,0)} (0)=f^{(1,0)}(\infty)=h^{(1,0)}(\infty)=1 \, .
\end{equation}
Thus, the asymptotic form of $H^{(1,0)}$ at $ r  \to \infty$ 
is
\begin{equation}
H^{(1,0)} \to v \exp\left[{\f{i\varphi}{2}}\right] \exp\left[{\f{i\varphi}{2} \sigma_3}\right] \, .\label{eq:asymptotic_10}
\end{equation}
From this form, it is clear that this string is a hybrid of halves of a global $U(1)_a$ string in Eq.~(\ref{eq:H-global}) 
and a non-topological $Z$-string in Eqs.~(\ref{eq:H-nontop}) and (\ref{eq:Z-nontop}).
The profile functions $f^{(1,0)}$, $h^{(1,0)}$ and $w^{(1,0)}$ are determined by solving the EOMs.
See Fig.~\ref{fig:10string} for the numerical solution.
Note that $h(r)$ does not necessarily vanish at the center of the string, $r=0$,
since it does not have a winding phase.
Instead, it tends to obtain a non-zero condensation.
In particular, its condensation value becomes larger 
when $m_{h^0}$ is larger than the other masses (right panel). This property will be important in later sections.

On the other hand, there is the other solution, called the $ (0,1) $-string:
\begin{align}
 H^{(0,1)} &= v  \begin{pmatrix}
	        h^{(0,1)} ( r ) &0 \\
               0 & f ^{(0,1)} ( r ) e^{i\varphi} 
	      \end{pmatrix},\\
 Z_i ^{(0,1)} &=  \f{ \cos \theta_\mr{W}}{g}  \f{ \epsilon_{3ij}x^j}{ r ^2} \left(1- w ^{(0,1)}( r )\right)\label{012222_17Mar19},
\end{align}
with the asymptotic form of the Higgs fields, 
\begin{equation}
H^{(0,1)} \to v \exp\left[{\f{i\varphi}{2}}\right] \exp\left[{\f{-i\varphi}{2} \sigma_3}\right]\label{eq:asymptotic_01} 
\end{equation}
for $ r  \to \infty$.
The boundary conditions for $f ^{(0,1)}$, $h^{(0,1)}$ and $ w ^{(0,1)}$ are the same as the $(1,0)$-string.

Note that, the two strings, $(1,0)$- and $(0,1)$ strings, are related to each other by the CP transformation Eq.~\eqref{eq:CP2}.
Since we are interested in the CP symmetric case, we have
\begin{equation}
 f ^{(1,0)}(r) = f ^{(0,1)}(r) \equiv f(r), 
\h{2em} h ^{(1,0)}(r) = h ^{(0,1)}(r) \equiv h(r), 
\end{equation}
\begin{equation}
 w ^{(1,0)}(r) =w ^{(0,1)}(r) \equiv w (r) \, ,
\end{equation} 
and the two strings have degenerate tensions (energy per unit length).\footnote{
When the $U(1)_Y$ gauge coupling is turned off, 
these two strings are continuously degenerated  
parametrized by ${\mathbb C}P^1$ moduli 
associated with spontaneously broken custodial symmetry in the vicinity of the string  
\cite{Eto:2018tnk,Eto:2018hhg}. 
Such a string is called a non-Abelian string,
see, e.g., Refs.~\cite{Tong:2005un,Eto:2006pg,Shifman:2007ce,Shifman:2009zz,Eto:2013hoa} as a review. 
Such a continuous degeneracy is lifted by the $U(1)_Y$ gauge coupling 
that explicitly breaks the custodial symmetry, 
leaving the (1,0) and (0,1) strings as strings with the minimum tension.
}

Looking at the asymptotic forms in Eqs.~\eqref{eq:asymptotic_10} and \eqref{eq:asymptotic_01},
it is clear that 
both the $(1,0)$- and $(0,1)$-strings have winding number $1/2$ for the global $U(1)_a$ symmetry,
and thus they are topological vortex strings of the global type. 
This is easily seen by calculating the following topological current:\footnote{This normalization of $\mathcal{A}_i$ is different from that in Ref.~\cite{Eto:2020hjb} by a factor $8\pi$.}
\begin{equation}
 \mathcal{A}_i \equiv \f{1}{8\pi v^2}\epsilon_{ijk}\partial^j J^k\label{eq:topological-charge} \, ,
\end{equation}
where $J_i$ is the Noether current of the $U(1)_a$ symmetry and defined as
\begin{equation}
 J_i = -i\, \mathrm{Tr}\left[H^\dagger D_i H - (D_i H)^\dagger H\right].
\end{equation}
This current is topologically conserved, $\partial_i \mathcal{A}_i=0$.
For the $(1,0)$- and $(0,1)$-strings located along the $z$ axis, 
$\mathcal{A}_i $ is obtained as 
\begin{align}
  \mathcal{A}_3 &= \f{1}{8\pi v^2}\left[\partial^1J^2 - \partial^2 J^1\right] \nonumber \\
&=\f{1}{8\pi r}\partial_{r}\left[2 f(r)^2-(1-w(r))(f(r)^2 -h(r)^2)\right],\label{eq:topcharge-strings}
\end{align}
which is half-quantized after integrating over the $x-y$ plane,
\begin{equation}
q_a\equiv 2\pi \int_0 ^\infty dr r \mathcal{A}_3  = \f{1}{2} \, ,
\end{equation}
and the other components vanish, $\mathcal{A}_1=\mathcal{A}_2=0$.
Therefore, the two strings are topologically stable
and cannot be removed by any continuous deformation.

Similarly to standard global vortices, their tensions (masses per unit length)
logarithmically diverge. It can be seen
from the kinetic term of the Higgs field:
\begin{equation}
 2\pi \int d r   r  ~\mr{Tr}|D_i H^{(1,0)}|^2
  \sim 2\pi \int d r   r  ~\mr{Tr}|D_i H^{(0,1)}|^2
  \sim \pi v^2 \int \f{d r } { r } \label{202812_20Dec19}
\end{equation}
for $ r \to \infty$.
This is a quarter of that for a singly quantized global $U(1)_a$ vortex
because of the half winding number for $U(1)_a$ \cite{Eto:2018tnk}.

On the other hand,
they also have a winding number $1/2$ inside the gauge orbit $U(1)_Z\subset SU(2)_W\times U(1)_Y$,
which lead to magnetic $Z$ fluxes inside them 
as magnetic flux tubes.
The amounts of the fluxes of $(1,0)$- and $(0,1)$-string are
\begin{equation}
 \Phi_Z^{(1,0)}= \frac{2 \pi \cos \theta_\mr{W} }{ g}, \h{2em}  \Phi_Z^{(0,1)}= - \frac{2 \pi  \cos \theta_\mr{W} }{ g}, 
\label{eq:Z_fluxes}
\end{equation}
along the $z$-axis, respectively.
They are half of that of the non-topological $Z$-string (Eq.~\eqref{eq:Zflux-nontop}) because of the half winding number.
Note that the amounts of the $Z$ fluxes for the $(1,0)$- and $(0,1)$-strings are different for generic $\tan \beta$, 
but are exactly same for our case $\tan \beta =1$, in contrast to 
the logarithmic divergent energy common for the both strings. 
At large distances from the strings, the $Z$ flux decays exponentially fast as a usual ANO vortex 
\cite{Abrikosov:1956sx,Nielsen:1973cs} in the Abelian-Higgs model, in contrast to the $1/ r $ tail given in Eq.~(\ref{202812_20Dec19}).
In other words, contributions to the energy from the non-Abelian gauge parts do not diverge.

In the presence of the $U(1)_a$ symmetry breaking terms, 
the asymptotic potential of the ansatz (\ref{eq:H(1,0)}),
in which the azimuthal angle $\varphi$ is replaced with a monototicaly increasing function 
$\phi(\varphi)$ of it 
 with the same range ($\phi(0)=0$ and $\phi(2\pi)=2\pi$),
becomes 
\begin{align}
V &\sim (-m_3^2 + 2 \alpha_6 v^2) \det H^{(1,0)}  
+ \alpha_5 (\det H^{(1,0)} )^2 + {\rm h.c.} \nonumber\\
&=  2(-m_3^2 + 2 \alpha_6 v^2) v^2 \cos \phi + 2\alpha_5 v^2 \cos 2\phi. 
\label{eq:double-SG} 
\end{align}
This is a double sine-Gordon potential, and thus a single $(1,0)$ topological $Z$ string is attached by 
at most two domain walls \cite{Eto:2018hhg,Eto:2018tnk}. 
Depending on the parameters, one or two walls are attached to the string.
The same holds for the $(0,1)$ string.
Compared with the asymptotic potential in Eq.~(\ref{eq:quad-SG}) for a single $U(1)_a$ string, 
the $U(1)_a$ string is found to be split into two fractional $Z$-strings $(1,0)$ and $(0,1)$ 
with being pulled by domain walls.
Below, we again assume the $U(1)_a$ symmetry.

\subsection{Asymptotics of topological $Z$-strings}
We here review the asymptotic forms of the $Z$-strings at large distances given in Ref.~\cite{Eto:2020hjb}.
For general local vortices, e.g., the ANO vortices in the Abelian-Higgs model or superconductors,
an asymptotic form is given by an exponentially damping tail whose typical size is the mass scale of the model.
The stability of a vortex lattice structure of the ANO vortices (called as an Abrikosov lattice) is determined
by a ratio between sizes of tales of the scalar (Higgs) and gauge fields,
which is equal to the ratio of the scalar and gauge couplings.
On the other hand, for global vortices (e.g., axion strings),
the asymptotic form is given by a power-law tail because of the massless NG boson (axion particle).
This means that global vortices are much fatter than local ones
and that they have logarithmically divergent tensions.
In the present case for the 2HDM,
there are various mass scales in the mass spectrum as shown in Sec.~\ref{sec:model},
so that the asymptotic form of the electroweak strings are quite non-trivial.
This situation is quite similar to non-Abelian vortices in dense QCD \cite{Balachandran:2005ev,Nakano:2007dr,Nakano:2008dc,Eto:2009kg,Eto:2009bh,Eto:2009tr}, 
see Ref.~\cite{Eto:2013hoa} as a review.


Let us consider the $(1,0)$-string.
By introducing new functions, the expression \eqref{eq:H(1,0)} can be rewritten as
\begin{align}
 H^{(1,0)}= \f{1}{2}v e^{i\varphi /2} ~e^{i\varphi \sigma_3/2 } \left(F( r ) \bm{1} + G( r )\sigma_3\right),
\end{align}
where
\begin{equation}
 F( r )\equiv f( r )+ h( r ), \h{2em} G( r ) \equiv f( r )-h( r ).\label{eq:new-profile-function}
\end{equation}
Here, $F$ and $G$ are profile functions in the mass basis.
The former corresponds to the custodial singlet component $\chi^0$ $(h^0)$ in Eq.~\eqref{010632_18Mar20}
and the latter is the $\sigma_3$ component of the (split) custodial triplet, $\chi_3$ ($A^0$).
We study the asymptotic forms of $F$, $G$ and $w^{(1,0)}$
at large distances compared to the inverses of the mass scales.
In this region, they are almost in the vacuum,
so that it is convenient to expand them around the vacuum as
\begin{eqnarray}
 && F( r )= F(\infty) + \delta F( r ) = 2+ \delta F( r ),\\
 && G( r ) = G(\infty) + \delta G( r ) = \delta G( r ),\\
 && w( r ) = w(\infty) +\delta w( r ) = \delta w( r ). 
\end{eqnarray}
For $m_Z < m_{A^0}$, after solving linearized EOMs, we obtain \cite{Eto:2020hjb}
\begin{align}
 \delta F &\simeq - \frac{1}{2(m_{h^0})^2r^2} + q_F \sqrt{\f{\pi}{2(m_{h^0})r}} e^{-(m_{h^0})r} + \mathcal{O}(r^{-4}) \label{eq:asymp1_F} \\
 \delta G( r ) &\simeq q_Z \f{1}{\left(m_Z^2 - (m_{A^0})^2\right) r ^2}~\sqrt{\f{\pi m_Z  r }{2}} e^{ -m_Z  r },\label{eq:asymp1_G} \\
 \delta w &\simeq q_Z \sqrt{\f{\pi m_Z r }{2}} e^{ -m_Z  r }  \label{eq:asymp1_w} 
\end{align}
where $q_F$ and $q_Z$ are integration constants which can be 
determined only by a numerical computation solving the EOMs.
Note that the leading term in $\delta F$ is the polynomial form $1/ (m_{h^0} r)^2$ due to the $U(1)_a$ NG mode
while $ \delta G$ and $\delta w$ have the same exponential tails $e^{-m_Z  r }$.

On the other hand, for $m_Z > m_{A^0}$, the asymptotic form of $\delta G( r )$ and $ \delta w( r )$ change as
\begin{align}
 \delta G(r) &\simeq q_G \sqrt{\f{\pi }{2 (m_{A^0})r}} e^{ -(m_{A^0})  r } \label{eq:asymp2_G} \, , \\
 \delta w( r ) &\simeq q_G \f{m_Z^2}{ (m_{A^0})^2 - m_Z^2}~\sqrt{\f{\pi}{2 (m_{A^0})r}} e^{ - (m_{A^0})  r }\label{eq:asymp2_w} \, ,
\end{align}where $q_G$ is an integration constant 
and 
$\delta F$ is the same as Eq.~\eqref{eq:asymp1_F}.

Therefore, the exponential tails of $\delta G$ and $\delta w$ are given by the lighter mass of between $m_Z$ and $m_{A^0}$
This is similar to a non-Abelian vortex in dense QCD~\cite{Eto:2009kg}.
In the both cases, the integration constants cannot be determined by the present argument.
To determine it, one has to solve the EOMs numerically and fit them via the above asymptotic forms.
The determined values are always $\mathcal{O}(1)$ values~\cite{Eto:2020hjb}.


\section{Interaction between topological $Z$-strings}
\label{sec:interaction}

\begin{figure}[tbp]
\centering
\includegraphics[width=0.7\textwidth]{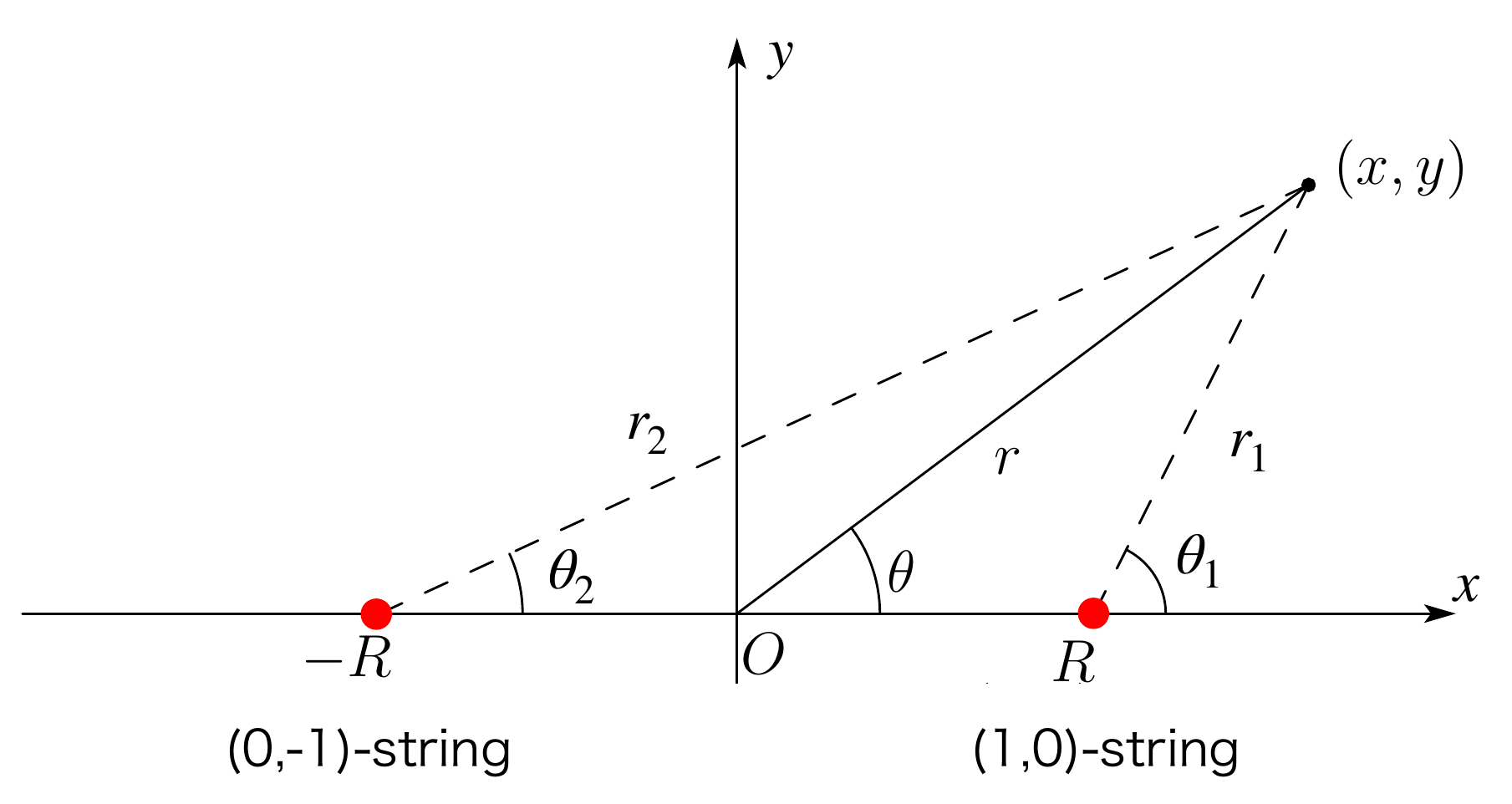}
\caption{
Two separated topological $Z$-strings, $(1,0)$ and $(0,-1)$ strings
located at $(x,y)=(\pm R ,0)$ on the $xy$ plane, respectively.
The strings are represented by two red dots.
}
\label{fig:two-strings}
\end{figure}

In this section, we discuss the interaction between 
two well-separated parallel topological $Z$-strings, 
extending along the $z$-axis. 
We consider a particular set in this section; 
one is a $(1,0)$-string in Eq.~\eqref{eq:H(1,0)}, 
and the other is a $(0,-1)$ string 
obtained by flipping the upside and downside of 
a $(0,1)$-string ($180^\circ$ rotation around the $x$ axis):
\begin{align}
 H^{(0,-1)} &= v  \begin{pmatrix}
	        h ( r ) &0 \\
               0 & f  ( r ) e^{-i\varphi} 
	      \end{pmatrix},\label{eq:H(0,-1)}\\
 Z_i ^{(0,-1)} &=  -\f{ \cos \theta_\mr{W}}{g}  \f{ \epsilon_{3ij}x^j}{ r ^2} \left(1- w ( r )\right) \label{eq:Z(0,-1)}.
\end{align}
The $(0,-1)$-string has a $U(1)_a$ winding phase opposite to that of the $(1,0)$-string
while it has the same $Z$ flux as the $(1,0)$-string.
Thus, a pair of the $(1,0)$- and $(0,-1)$-strings is topologically trivial,
and it is identical to a single non-topological $Z$-string
when it is viewed from a far distance.
The cross section of the two strings on the $x$-$y$ plane 
are point-like. 
We assume that they are separated with a distance $2R$, see Fig.~\ref{fig:two-strings}.

In this section, we study how the two strings interact with each other.
We assume the exact (and spontaneously broken) $U(1)_a$ symmetry in the Higgs potential 
in which case no domain walls are attached to the strings.
As stated in Sec.~\ref{sec:model}, to make the NG boson massive,
this symmetry must be explicitly broken by turning on the $U(1)_a$-breaking terms in the Higgs potential.
The effects of the explicit breaking terms will be discussed later.
Below, we will show that 
they feel the following repulsive or attractive forces:
\begin{description}
 \item[(a)] long-range attractive force from the global winding number of $U(1)_a$ symmetry
 \item[(b)] short-range attractive force by the custodial singlet $\chi^0$ (SM-like Higgs boson $h^0$)
 \item[(c)] short-range repulsive force due to the $Z$ fluxes
 \item[(d)] short-range repulsive force from the condensation of $h(r)$ when $m_{h^0}\gg m_{A^0}$.
\end{description}
These forces have typical reach distances:
(a) infinity, (b) $1/m_{h^0}$, (c) $1/m_Z$ and (d) vortex core width $\xi$.

In order to show the first three ones, (a)-(c), 
it is sufficient to use the asymptotic formulae presented in the above
by assuming that they are sufficiently separated.
We thus use the following approximated ansatz to describe the two-vortex system:
\begin{eqnarray}
 H &\simeq& \frac{1}{v} H^{(1,0)} \cdot H^{(0,-1)}\\
 \vec{Z} &\simeq& \vec{Z}^{(1,0)} + \vec{Z}^{(0,-1)}
\end{eqnarray}
where the dot `` $\cdot$ '' for $H$ represents a product for each component in the matrices.
Namely, we consider
\begin{eqnarray}
 H &\simeq& v 
\begin{pmatrix}
f(r_1) h(r_2) e^{i \theta_1} & 0 \\
0 & f(r_2) h(r_1) e^{-i \theta_2}
\end{pmatrix}\label{170642_15Jun21} \\
 \vec{Z} &\simeq& \f{\vec{e}_{\theta_1}}{r_1} \frac{1}{g_Z}(1-w(r_1)) + \f{\vec{e}_{\theta_2}}{r_2} \frac{1}{g_Z}(1-w(r_2)),\label{201102_22Jun21}
\end{eqnarray}
where $r_1$, $r_2$, $\theta_1$, and $\theta_2$ are defined as Fig.~\ref{fig:two-strings},
and $\vec{e}_{\theta_i}$'s denote unit vectors in the angular direction around the strings.

The tension, \textit{i.e.}, the total energy per unit length integrated over the $x$-$y$ plane, 
in the presence of the two strings is defined as
\begin{equation}
 T \equiv \int d^2 x \, (K_H + K_Z + V) \, ,
\end{equation}
where
$K_H$ and $K_Z$ are the kinetic energy of the Higgs field and the $Z$ gauge fields, given respectively by 
\begin{eqnarray}
 K_H &\equiv& \mr{Tr}|D_i H|^2 \,  ,\\
  K_Z &\equiv& \f{1}{2} (Z_{ij})^2 = \f{1}{2} (\vec{\nabla} \times \vec{Z})^2 \, ,
\end{eqnarray}
and $V$ is the potential energy.
Since we are interested in the interaction of the strings,
we consider the difference of the tension from those of single $(1,0)$ and $(0,-1)$ strings,
\begin{equation}
  \delta T \equiv T - T^{(1,0)} -T^{(0,-1)} \, ,\label{eq:tension-diff}
\end{equation}
where $T^{(1,0)}$ and $T^{(0,-1)}$ are the tensions of $(1,0)$ and $(0,-1)$ strings, respectively,
and the interaction force between the strings can be calculated by the following definition
\begin{equation}
 F (R)\equiv - \f{d}{dR} \delta T \, .
\end{equation}

It is rather complicated to present the full expression of $\delta T$.
Instead, we just demonstrate how 
(a) the global winding number, (b) the custodial singlet $\chi^0$, and (c) the $Z$ fluxes contribute to $F(R)$.
For the detailed computation, see Appendix \ref{app:interaction}.
Let us first concentrate on  (a), \textit{i.e.}, the contributions from the global winding phases.
It comes from the gradient energy of the winding phases, $|\partial_i e^{i\theta_1}|^2$ and $|\partial_i e^{i \theta_2}|^2$, of the Higgs fields 
(only half of which is canceled by the $Z$ gauge field)
and is associated with the global winding number of $U(1)_a$ symmetry,
which is similar to the ordinary long-range force between integer global vortices like axion strings.\footnote{More precisely, this situation may be similar to 
interaction between two vortices in two-component Bose-Einstein condensates \cite{Eto:2011wp}.
}
Substituting the asymptotic expressions in Eqs.~\eqref{eq:asymp1_F}-\eqref{eq:asymp2_G} and using Eq.~\eqref{eq:new-profile-function},
it is found (see App.~\ref{app:interaction}) that such a part has $\mathcal{O}(r^{-2})$ forms as
\begin{align}
  \delta T \big|_\text{(a)}  &
\simeq \int d^2x ~ v^2  \left[
- \frac{\cos (\theta_1 - \theta_2)}{r_1 r_2} 
+ \mathcal{O}(r^{-4},\delta F, \delta G, \delta w) \right]\, ,\label{eq:deltaT-a}
\end{align}
where we have omitted sub-leading terms $\mathcal{O}(r^{-4})$
since the strings are sufficiently separated.
This integration over the $x$-$y$ plane can be performed analytically, to give
\begin{align}
\delta T\big|_\text{(a)} & =  2 \pi v^2 \log \f{R}{\xi}  \, . \label{eq:deltaT-global}
\end{align}
We have introduced the UV cutoff $\xi$, which is given by a width of the string core 
(typically, inverse of the mass scale of the lightest massive particle).
This logarithmic behavior is the same as that of ordinary global vortices.
Thus, a contribution to the asymptotic force between the vortices can be calculated as
\begin{align}
\left. F(R) \right|_{\text{(a)}}
 =& - \frac{2\pi v^2}{R} \, ,\label{eq:force(a)}
\end{align}
where the negative sign means that it is an attractive force.
Therefore, the global winding number gives the attractive interaction.
Note that this interaction is a long-range force 
because Eq.~\eqref{eq:force(a)} has the same power law as the two-dimensional Coulomb force,
corresponding to the exchange of the $U(1)_a$ NG boson $H^0$.
This expression is reliable only when the strings are not overlapped.

Next, 
let us consider (b) the attractive force mediated by the custodial singlet $\chi^0$.
Since the custodial singlet $\chi^0$ is described by the profile function $F$,
we concentrate on terms containing $F(r)$ in the tension.
Again, substituting the asymptotic expressions in Eqs.~\eqref{eq:asymp1_F}-\eqref{eq:asymp2_G} into the tension in Eq.~\eqref{eq:tension-diff},
it is found that its leading contribution behaves as $\mathcal{O}(r^{-4})$ in the integrand of the tension.
This can be seen by ignoring contributions from the other components as (see App.~\ref{app:interaction})
\begin{align}
  \delta T \big|_\text{(b)}  &\simeq \int d^2x ~ v^2  \left[
  \frac{1}{2(m_{h^0})^2} \frac{\cos (\theta_1 - \theta_2)}{r_1 r_2} \left(\frac{1}{r_1^2} + \frac{1}{r_2^2}\right)
 - \frac{1}{4(m_{h^0})^2 r_1^2r_2^2} \right] \n \\
&+ \mathcal{O}(r^{-6},\delta G, \delta w) \, . \label{eq:deltaT-b}
\end{align}
The integration over the $x$-$y$ plane can be performed analytically, to yield
\begin{align}
\delta T \big | _\text{(b)}& =  
 \frac{\pi v^2}{2(m_{h^0})^2 R^2} - \frac{\pi v^2}{4(m_{h^0})^2 R^2} \log \f{R}{\xi} +
\text{const.} \, ,\label{eq:deltaT-singlet}
\end{align}
and thus asymptotic force between the strings is calculated as
\begin{align}
\left. F(R) \right|_{\text{(b)}}
 =& - \frac{\pi v^2}{R} \left[\frac{1}{2(m_{h^0})^2 R^2}\left( \log \f{R}{\xi} - \frac{5}{2}\right)\right] \, ,\label{eq:force(b)}
\end{align}
where the terms in the square bracket is positive for sufficiently large $R$
and provides the attractive force between the strings.\footnote{This force is in fact the same with  
that between two vortices $(1,0)$ and $(0,1)$ in two-component Bose-Einstein condensates \cite{Eto:2011wp}.
}
Therefore, the custodial singlet component gives the attractive interaction as well as the global winding number (a).
Unlike the long-range force (a), this is a short-range force because it damps for a sufficiently large $R$ compared to the length scale $(m_{h^0})^{-1}$.

Let us consider (c),
\textit{i.e.}, the $Z$ fluxes giving a repulsive force by the gauge kinetic energy $K_Z$.
To see this, we concentrate on $K_Z$ in $\delta T$ as
\begin{align}
 \delta T\big|_\text{(c)}
=& \int d^2x \, K_Z \nonumber \\
=& \int d^2x  \, \f{1}{g_Z^2} \f{\delta w ' (r_1) \delta w'(r_2)}{r_1 r_2} \nonumber \\
=& \int d^2x   \f{\pi q_Z^2  v^2  m_Z }{8\sqrt{r_1 r_2}}\, e^{ -m_Z (r_1 + r_2) }  
\end{align}
where we have assumed that $m_Z$ is lighter than $m_{A^0}$ 
and used the asymptotic expression in Eq.~\eqref{eq:asymp1_w}.
This integration can be rewritten in terms of dimensionless variables as
\begin{align}
& q_Z^2 v^2 \frac{\pi m_Z}{8} \int d^2 x \frac{1}{\sqrt{r_1 r_2}}  
 e^{-m_Z (r_1 + r_2)} \nonumber \\[1ex]
=& \, q_Z^2 v^2 \frac{\pi m_Z}{8} \int d^2 x 
\left[r^4 + R^4 - 2 r^2 R^2 \cos{2\theta}\right] ^{-1/4} \nonumber \\[1ex]
& \h{3em} \times\exp\left[-m_Z \left(\sqrt{r^2+R^2 - 2 rR \cos \theta} + \sqrt{r^2+R^2 + 2 rR \cos \theta}\right)\right] \nonumber \\[1ex]
= & \,q_Z^2 v^2 \frac{\pi m_Z R}{8} \int_0^\infty dt\, t \int d \theta 
 \left[ t^4 + 1 - 2 t^2 \cos{2\theta}\right] ^{-1/4} \nonumber \\[1ex]
& \h{3em} \times\exp\left[- m_Z  R \left(\sqrt{t^2+1 - 2 t \cos \theta} + \sqrt{t^2+1+ 2 t \cos \theta}\right)\right] \h{1em} (t\equiv r/R) \nonumber \\[1ex]
\equiv & \, \f{q_Z^2 v^2 \pi}{8} y \mathcal{F}(y) \h{3em} (y \equiv R \, m_Z) \, .
\end{align}
The integration over $t$ and $\theta$ cannot be performed analytically.
The interaction force produced by $K_Z$ is expressed as
\begin{align}
 \left.  F(R) \right|_\text{(c)} &=  - m_Z \f{q_Z^2 v^2 \pi}{8} \f{d}{d y }  y \mathcal{F}(y) .
\end{align}
The quantity $(y \mathcal{F}(y))'$ can be obtained numerically, as shown in Fig.~\ref{fig:Dyf}.
It is clear that it is always positive for large $y$ (or $R$), giving the repulsive interaction (c).
Again, this is a short-range force 
because it damps for a large $R$ compared to the length scale $(m_Z)^{-1}$.

\begin{figure}[tbp]
\centering
\includegraphics[width=0.5\textwidth]{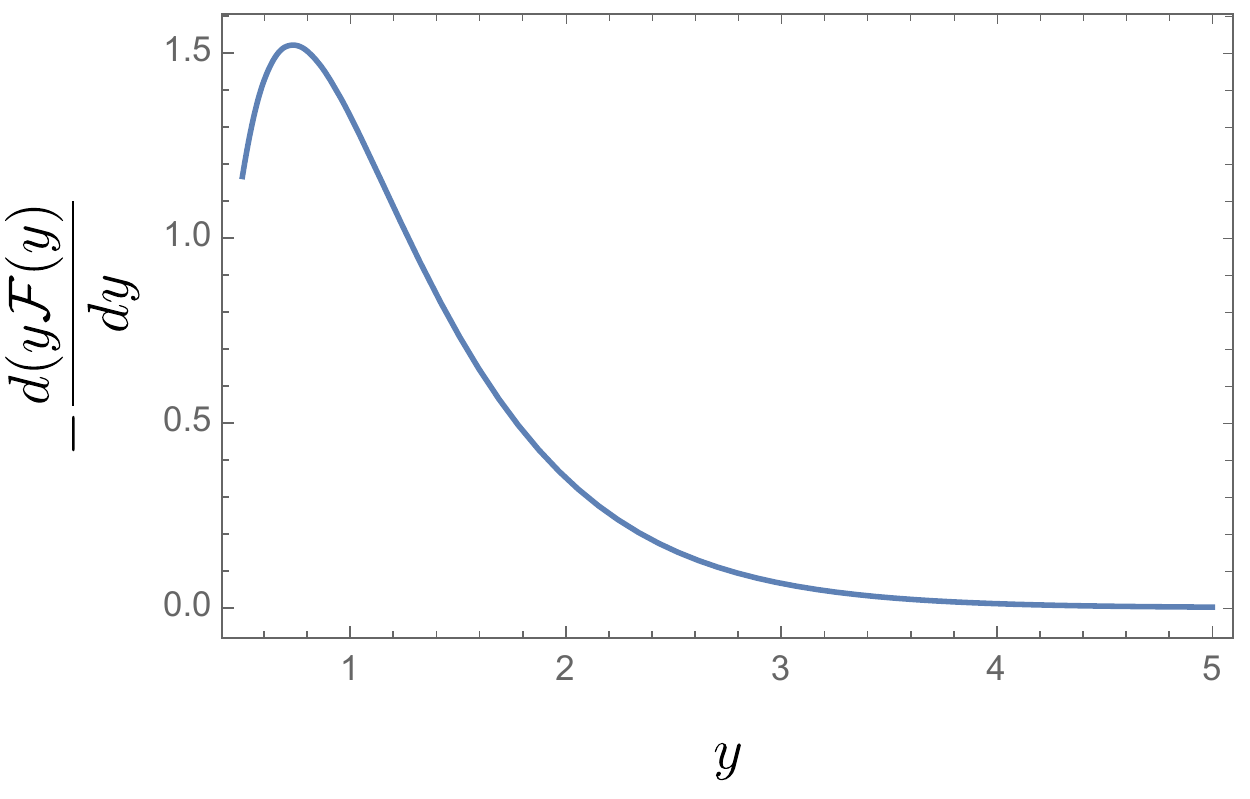}
\caption{
Plot of the function $-(y \mathcal{F}(y))'$.
It is positive for arbitrary $y > 0$.
}
\label{fig:Dyf}
\end{figure}

Finally, we show (d),
\textit{i.e.}, that the repulsive force arises due to the condensation of the field $h(r)$
when $m_{h^0}\gg m_{A^0}$.
In this case, we cannot use the asymptotic expression for the strings 
because this force is significant only when the two tails of the profile function $h(r)$ are overlapped.
For $(1,0)$-string (Eq.~\eqref{eq:H(1,0)}), the lower-right component of the matrix $H$, $h(r)$, 
condensates at the center.
In particular, this condensation value becomes large when $m_{h^0}\gg m_{A^0}$, as stated above
and shown in the right panel of Fig.~\ref{fig:10string}.
On the other hand, for $(0,-1)$-string (Eq.~\eqref{eq:H(0,-1)}),
the lower-right component is described by the profile function $f(r)$, which has the winding phase 
and must vanish at its center.
(Inversely, the upper-left component vanishes for $(1,0)$-string but does not for $(0,-1)$-string.)
For the continuity, the lower-right component has to change from a non-zero condensation to 0 with the length scale $R$ which we assume to be of order of the width of the strings.
Therefore, when $(1,0)$- and $(0,-1)$-strings are close to each other,
the Higgs field changes rapidly and produces a kinetic energy,
which is roughly estimated as
\begin{align}
 K_H & \sim \int dx dy \, v^2 \left(\partial_x h(r)\right)^2 \nonumber\\
& \sim v^2 \xi R \left(h(0)/R \right)^2 \nonumber \\
& \sim v^2 h(0)^2 \xi/R
\end{align}
where the integration of $x$ and $y$ give the factors $R$ and $\xi$, respectively.
Thus the interaction force is estimated as
\begin{equation}
 F(R) \big|_\text{(d)}= - \frac{d}{dR}  v^2 h(0)^2 \f{\xi}{R} =  \f{v^2 h(0)^2 \xi}{R^2} >0 \, ,
\quad (\text{for } R\sim \xi)\, ,
\label{eq:F_d}
\end{equation}
which means that it gives the repulsive force.
As stated above, this is active only when the strings are close to each other.
Thus it is a short-range force and has a typical length scale of the width of the string core.

We have shown that the two strings, $(1,0)$ and $(0,-1)$ strings, feel the interaction forces (a), (b), (c) and (d).
Which force is dominant highly depends on the model parameters, and to investigate it, 
one needs to perform more complicated arguments,
which seem almost impossible in the analytic approach in general.

Before closing this section, we discuss the effect of the $U(1)_a$-breaking terms in the Higgs potential.
Due to the explicit breaking terms,
the phase directions of the $(1,0)$ and $(0,-1)$ strings feel the sine-Gordon potential Eq.~\eqref{eq:double-SG},
which deforms the tension $\delta T|_\text{(a)}$.
As a result, there appear domain walls stretching between the two strings as the axion strings attached with the axion domain walls.
This means that the strings feel a constant attractive force due to the wall tension, 
instead of the Coulomb-like force $F(R)|_\text{(a)}$ arising from the massless NG boson.
Namely, the force (a) is replaced by (a)':
\begin{description}
 \item[(a)'] confining attractive force by the $U(1)_a$ domain wall
\end{description}
However, as long as the $U(1)_a$-breaking parameters are sufficiently small, 
or the string distance is sufficiently smaller than the domain wall width,
the wall tension is not significant, and hence the above picture and analysis are qualitatively correct.

\section{Analytic argument on vortex molecule}
\label{sec:analytic-molecule}
We now consider a bound state of the two $Z$-strings, $(1,0)$ and $(0,-1)$ strings,
in the case with the exact (and spontaneously broken) $U(1)_a$ symmetry.
We call this bound state \textit{vortex molecule}.
Note that, this configuration is in the same topological sector as that of the non-topological $Z$-string Eqs.~\eqref{eq:H-nontop} and \eqref{eq:Z-nontop}.
Indeed, if this configuration shrinks into a single vortex-like configuration,
it has the same $Z$ flux as that of the non-topological $Z$-string Eqs.~\eqref{eq:H-nontop} and \eqref{eq:Z-nontop}
and no $U(1)_a$ topological charge as mentioned above.

It is known \cite{James:1992zp,Goodband:1995he} that the non-topological $Z$-string is unstable.
It hence seems that the vortex molecule is also unstable due to the same reason.
However, this is not the case.
This can be understood intuitively as follows.
Let us consider the $(1,0)$ and $(0,-1)$ strings distanced with a length scale $L$ on the $xy$ plane.
For $L\to \infty$, 
this configuration is of course stable because each string has the non-zero topological charges $\pm 1/2$, respectively,
and they do not feel each other due to the infinite separation.
For finite but sufficiently large $L$, 
the configuration is still stable 
and each string does not decay as long as the strings do not have significant overlap.
It naively seems that this molecule tends to shrink ($L$ decreases) 
because of the long-range attractive force (a) coming from the topological $U(1)_a$ charges.
However, this shrinking (decreasing of $L$) can be stopped by balancing between the attractive force and the short-range repulsive force (c) or (d).
If $L$ stops at a sufficiently long length compared to the width of the strings 
(more precisely, the length scale of the density of $U(1)_a$ topological charge),
the molecule remains static and stable.
This configuration has the ``polarized'' $U(1)_a$ topological charges (from $0$ of the non-topological $Z$-string to $1/2$ plus $-1/2$ 
of the molecule), resulting in its stability.

In this section, we study conditions to realize the stability of the vortex molecule in an analytic way.
A more quantitative analysis based on numerical simulations will be presented in the next section.

\subsection{Conditions to avoid shrinking}
Let us consider the same configuration as that in Fig.~\ref{fig:two-strings},
\textit{i.e.}, $(1,0)$ and $(0,-1)$ strings that are separated by the distance $2R$.
They are located at $(x,y)=(\pm R,0)$ on the $xy$ plane, respectively.
We call the length between the two strings a polarization length, $2R$.
As stated above, 
this vortex molecule seems to shrink into the non-topological $Z$-string (the polarization length $2R$ approaches to $0$)
since there is an attractive and long-range interaction between the strings, 
which comes from the global $U(1)_a$ winding phase
and is nothing but the force (a) shown in the last section.
However, this shrinking can be stopped and the polarization length can be kept a non-zero value
because there are also repulsive forces (c) and (d), which are short-range forces.
If the attractive one is balanced with the repulsive ones at a certain distance, 
the molecule stops to shrink and the polarization length is kept non-zero.

Let us obtain conditions to avoid shrinking,
which is a necessary condition for the molecule to be a stable solution of the EOMs.
There are a long-range attractive force (a), short-range attractive force (b),
and two short-range repulsive forces (c) and (d).
Naively, it is realized when the repulsive forces (c) and (d) are stronger than the attractive force (b).
In this case, the strings feel the attractive force at long distance 
while they feel the repulsive force at short distance.
Thus the repulsive and attractive forces are balanced at a certain distance.

As we mentioned in Sec.~\ref{sec:interaction}, the strength of the attractive force (b) is characterized by 
the mass of the custodial singlet component $\chi^0$ (SM-like Higgs boson)
since the attractive force (b) is mediated by $\chi^0$.
Hence it becomes weaker when the mass of $\chi^0$ is sufficiently large compared to the other mass scales.
On the other hand, the repulsive force from the $Z$ fluxes, (c), 
is characterized by the $Z$ boson mass $m_Z$.
It becomes stronger for lighter $Z$ boson mass $m_Z$.
Furthermore, the repulsive force (d) comes from the condensation of $h(r)$ in each vortex.
The condensation becomes strong when $m_{h^0}\gg m_{A^0}$ (see Sec.~\ref{sec:top-Zstring}).
Therefore, the short-range repulsive forces (c) and (d) are stronger than the short-range attractive force (b)
when the following inequality is satisfied:
\begin{equation}
 m_{h^0}\gg m_{Z},\, m_{A^0}\, .\label{eq:balance-condition}
\end{equation}
When this condition is met, the attractive force (b) is negligible,
and the long-range attractive force (a) is expected to be balanced with the short-range repulsive ones (c) and (d)
with a non-zero polarization length.
Thus the vortex molecule avoids shrinking.

\subsection{Polarization of topological charge}
Even if the molecule is prevented from shrinking,
it does not necessarily mean that the molecule does not decay.
This is because the configuration is in the same topological sector as that of the non-topological $Z$-string,
and there is no topological reason that ensures its stability.
More concretely, it might decay by the same mechanisms as the non-topological $Z$-string, 
\textit{e.g.}, the condensations of the $W$ boson and the (would-be) NG boson in the Higgs field.
However, this instability can be avoided.
When the polarization length is kept a sufficiently larger length than the width of the strings,
\textit{i.e.}, when the strings are well separated,
each string must be topologically stable because they have an isolated topological charge $q_a \simeq \pm 1/2$ protected by the $U(1)_a$ symmetry.
Thus the vortex molecule is stabilized by polarizing the topological charge $0$ into $+1/2$ and $-1/2$ for each string.

In the following, let us look closer at this point of the stability based on an analytic argument.
We assume that the vortex molecule has a polarization length $R$
and that the repulsive and attractive forces are balanced.
Then we consider the density of the topological current associated with the $U(1)_a$ winding phase, which is defined by Eq.~\eqref{eq:topological-charge}.
Since the strings are separated, 
$\mathcal{A}_3 $ is given by a superposition of those of $(1,0)$ and $(0,-1)$ strings,
\begin{equation}
\mathcal{A}_3 |_{\text{molecule}} \simeq 
\mathcal{A}_3 |_{(1,0)\text{-string}} + \mathcal{A}_3 |_{(0,-1)\text{-string}} \, ,\label{eq:topcharge-molecule}
\end{equation}
in which each term has a peak at $(x,y)=(\pm R,0)$,
see Fig.~\ref{fig:topological-charge} for the schematic picture.
As Eq.~\eqref{eq:topcharge-strings}, 
they are calculated as
\begin{align}
\mathcal{A}_3\big |_{(1,0)\text{-string}} &=\f{1}{8\pi  r_1}\partial_{r_1}\left[2 f(r_1)^2-(1-w(r_1))(f(r_1)^2 -h(r_1)^2)\right] \\
\mathcal{A}_3\big |_{(0,-1)\text{-string}} &=\f{1}{8\pi r_2}\partial_{r_2}\left[2 f(r_2)^2-(1-w(r_2))(f(r_2)^2 -h(r_2)^2)\right],
\end{align}
which vanish as $r_1 \to \infty$ and $r_2\to \infty$.

When the strings are sufficiently separated
so that the polarization length $2R$ is sufficiently larger than the typical size of the decaying tails of $\mathcal{A}_3\big |_{(1,0)\text{-string}}$ and $\mathcal{A}_3 |_{(0,-1)\text{-string}}$,
the two terms in the right-hand side in Eq.~\eqref{eq:topcharge-molecule} do not overlap with each other.
This is equivalent to that they approach to 0 rapidly at $r_1\simeq R$ and $r_2 \simeq R$,
\begin{equation}
 f(R)\simeq 1, \h{2em} h(R)\simeq 1 \, ,\label{eq:R-condition}
\end{equation}
which states that the Higgs field should rapidly approach to the vacuum value $H= v \bm{1}_{2\times 2}$ within the length scale $R$.
When this condition Eq.~\eqref{eq:R-condition} is satisfied, 
the molecule configuration has two isolated topological charges
(the left panel in Fig.~\ref{fig:topological-charge}),
\begin{align}
  q_a &= 2\pi\int_0 ^\infty dr r  \mathcal{A}_3 |_{\text{molecule}} \nonumber \\
& \simeq  2\pi\int_0 ^R dr_1 r_1  \mathcal{A}_3\big |_{(1,0)\text{-string}} 
+ 2\pi\int_0 ^R dr_2 r_2 \mathcal{A}_3\big |_{(0,-1)\text{-string}} \, ,
\end{align}
which are carried by the two strings, respectively.
Each string cannot decay because each polarized topological charge must be conserved,
resulting in the static and stable molecule.
Thus the vortex molecule is stabilized due to the polarization of the $U(1)_a$ topological charges.

Let us rephrase the condition Eq.~\eqref{eq:R-condition} into a more practical condition.
We assume Eq.~\eqref{eq:balance-condition} preventing the molecule from shrinking.
Let us recall the asymptotic behaviors of the profile functions of $(1,0)$ and $(0,-1)$ strings,
shown by Eqs.~\eqref{eq:asymp1_F}, \eqref{eq:asymp1_G} (for $m_Z < m_{A^0}$), and \eqref{eq:asymp2_G} (for $m_Z > m_{A^0}$).
In particular, $\delta G=f-h$ is proportional to $(m_{A^0}^2-m_Z^2)^{-1}$
and suppressed by a factor $ (m_{A^0})^{-2}$ when $m_{A^0}$ is much larger than $m_Z$.
Hence both of $\delta F$ and $\delta G$ decay rapidly 
and the condition Eq.~\eqref{eq:R-condition} is easily realized.
On the other hand, when $m_{A^0}$ is much smaller than $m_Z$,
the rapidity of the decay of $\delta G$ becomes milder and the condition Eq.~\eqref{eq:R-condition} is unlikely to be realized.
Therefore, we obtain a rough condition to satisfy Eq.~\eqref{eq:R-condition},
\begin{equation}
\text{molecule stability condition :} \, m_{h^0} \gg m_{A^0} \gg m_Z \, . \label{eq:stable-condition}
\end{equation}
When this condition is satisfied, the vortex molecule is expected to be static and stable.
This argument will be confirmed by numerical calculations below.

\begin{figure}[tbp]
\centering
\includegraphics[width=0.8\textwidth]{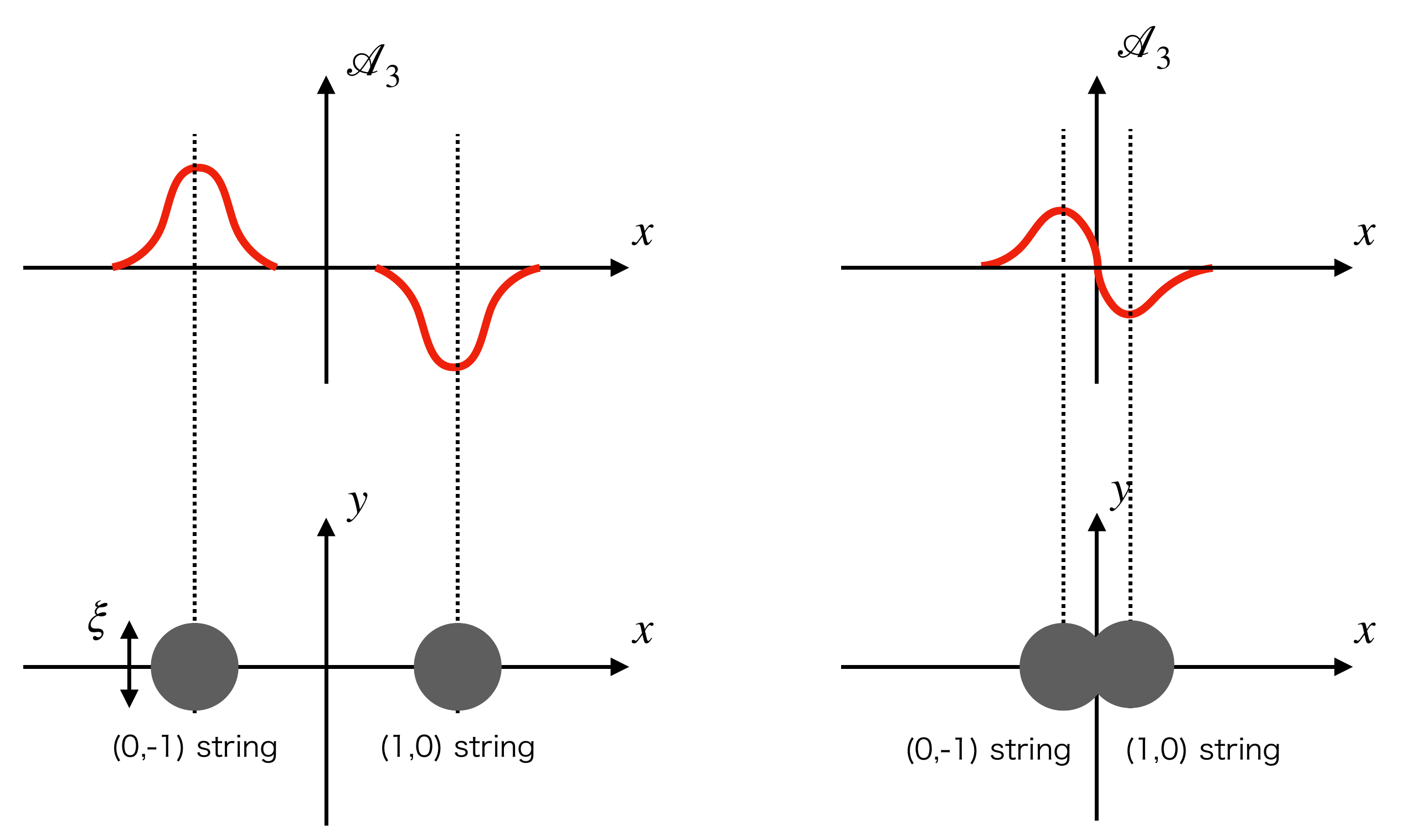}
\caption{
Schematic picture for the distribution of the topological current $\mathcal{A}_3$
for the $(1,0)$- and $(0,-1)$-strings (represented by the gray regions).
$\xi$ is the typical length scale of the string core.
The left panel shows the well-separated case,
in which $\mathcal{A}_3$ has negligible overlap,
and hence the vortex molecule consisting of the two strings 
does not decay as long as they are well-separated, namely, 
$R$ is sufficiently large compared with their core sizes.
The right panel shows the non-separated case,
in which $\mathcal{A}_3$ has finite overlap and its shape is deformed.
The molecule is not stable and decays as the non-topological $Z$-string.
}
\label{fig:topological-charge}
\end{figure}

\section{Numerical analysis}
\label{sec:numerical-molecule}
In this section, we present numerical analysis for the vortex molecule solution.
Firstly, we show 
two examples of the stable solution of the EOMs: one for the
$U(1)_a$ symmetric case and the other for the $U(1)_a$-broken case.
Then we study a parameter space in which the stable vortex molecule exists
with our focus being put on strength of breaking of 
the $U(1)_a$ symmetry.
The custodial symmetry is always imposed by the condition Eq.~\eqref{eq:cond_2},
leading to $\tan \beta = 1,~ m_{H^\pm} = m_{A^0}$.

\subsection{Vortex molecule solution}

We here present a stable vortex molecule solution.
Our numerical scheme is the following.
Let us write the equations of motion (EOMs) as ${\rm EoM}[\phi(x)]=0$  
for the static fields represented by $\phi(x)$. For our problem,
$\phi(x)$ stands for all the fields including scalar
and gauge fields. Instead of trying to solve the EOMs,
we introduce the relaxation time $\tau$ and modify
the static EOM as
\begin{eqnarray}
{\rm EoM}[\phi(\tau,x)] = \frac{\partial \phi(\tau,x))}{\partial \tau}.
\label{eq:relax}
\end{eqnarray}
We start with an initial configuration $\phi(\tau=0,x)$ which
satisfies the correct boundary condition at spatial boundary,
and integrating the EOMs toward ``future'' by means of $\tau$.
As the relaxation time
evolves, the configuration changes from time to time. 
Since the right hand side of Eq.~(\ref{eq:relax})
is a sort of dispersion term of energy, the initial configuration
evolves toward convergence at a local energy minimum.
When we reach the convergence after a long simulation time,
we regard the final state as a stable solution of the original EOMs since 
the right hand side of Eq.~(\ref{eq:relax}) 
becomes zero (within a numerical accuracy).


Firstly, as a benchmark case, 
we impose the $U(1)_a$ symmetry (Eq.~\eqref{eq:cond_1}) leading to $ m_{H^0}=0$,
and take the remaining parameters as
\begin{equation}
  m_{h^0}=1000 \, \mr{GeV} , ~  v = v_\mr{EW}/2 = 123 \, \mr{GeV} \, , \nonumber
\end{equation}
\begin{equation}
m_{H^\pm}  (=m_{A^0})  =750 \, \mr{GeV}  \, ,  \nonumber
\end{equation}
\begin{equation}
  m_Z  =91 \, \mr{GeV} , ~\sin^2 \theta_W = 0.23 \, .\label{eq:parameters1}
\end{equation}
Fig.~\ref{Fig:molecule} shows plots of the obtained solution.
The upper-left, upper-right, and the lower panels show
the profile of $\mr{Tr} |H|^2$, the $Z$ flux, and the energy density, respectively.
In addition, Fig.~\ref{Fig:molecule2} shows plots of $\mr{det} |H|$.
The gauge invariant quantity $\mr{det} |H|$ is useful to measure the positions of the strings.
In the figure, it goes to zero at two points on the $xy$ plane, $(x,y)\simeq (\pm 0.6,0)$,
which means that there are two separated strings.
It is also clear that the configuration contains the $Z$ flux as in the case of the non-topological $Z$-string.
The flux does not have rotational symmetry in contrast to the non-topological $Z$-string.
The total $Z$ flux on the $xy$ plane is calculated as $\simeq 4 \pi /g_Z$.
The energy density has two peaks corresponding to the two strings.

\begin{figure}[tbp]
 \centering
\includegraphics[width=0.47\textwidth]{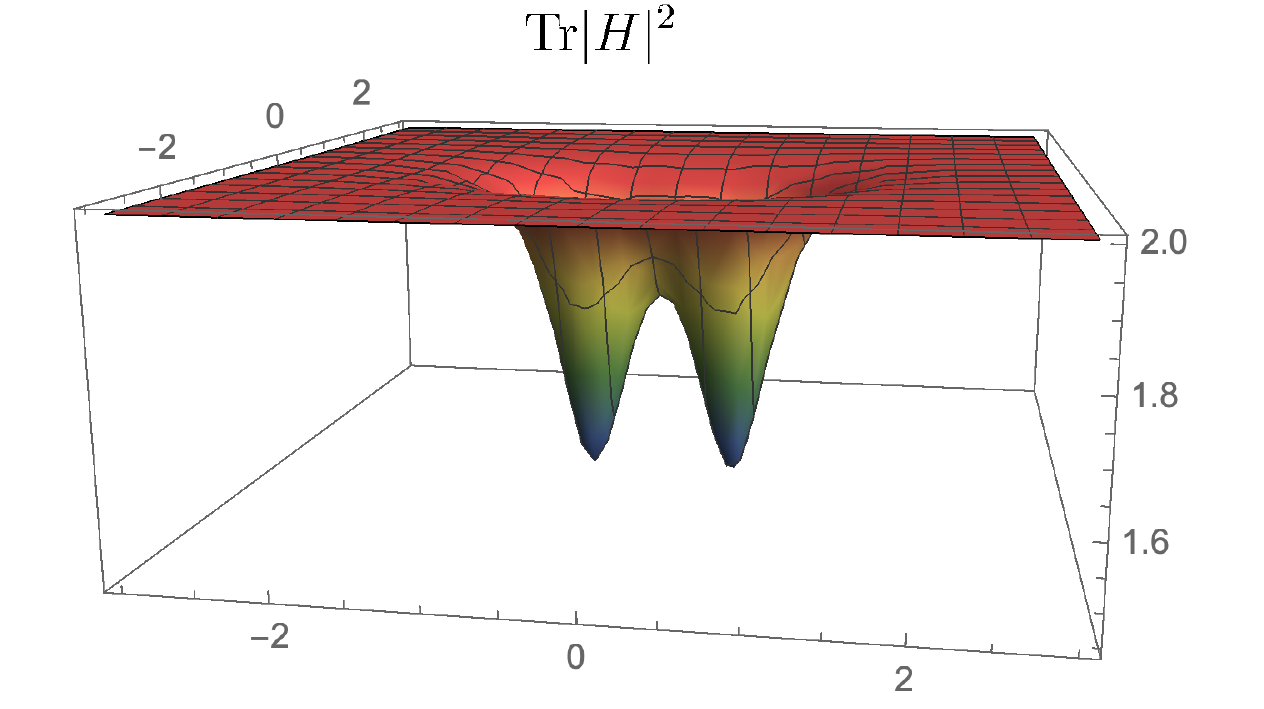} \hspace{2em}
\includegraphics[width=0.43\textwidth]{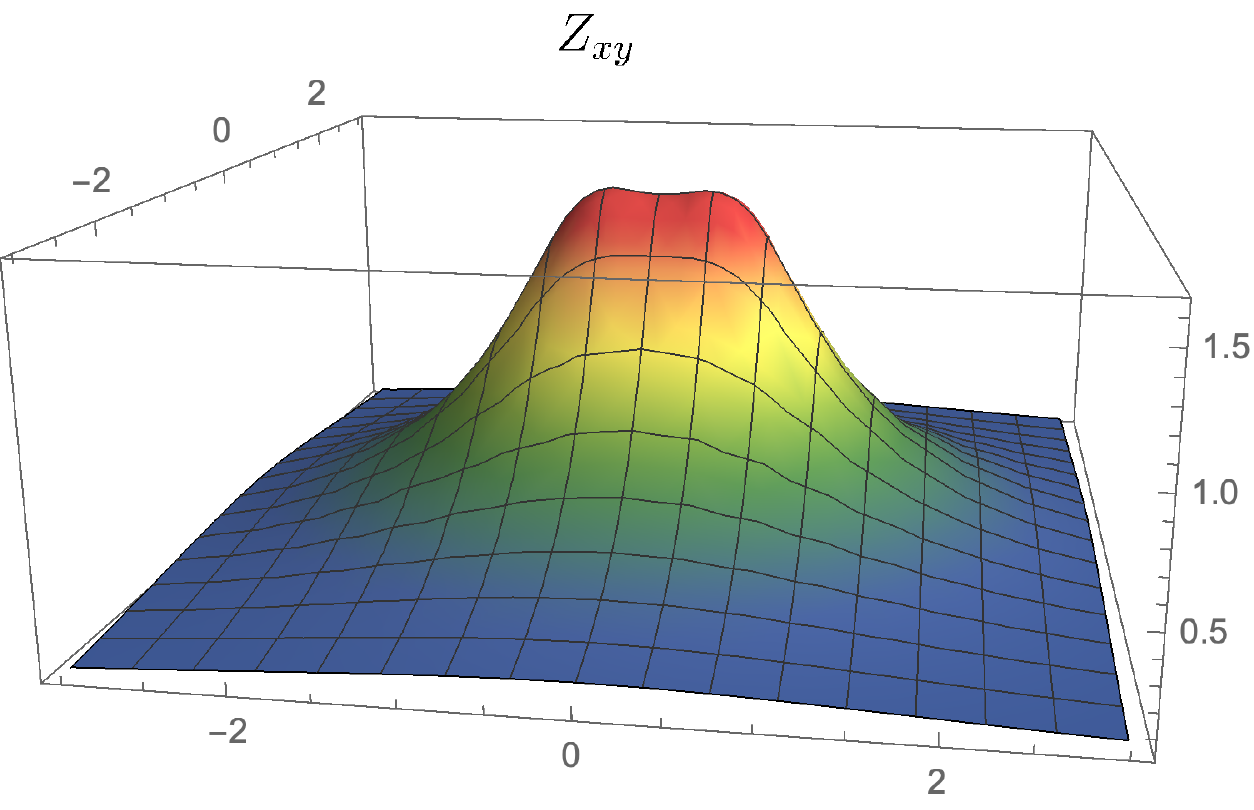} \\[5ex]
\includegraphics[width=0.45\textwidth]{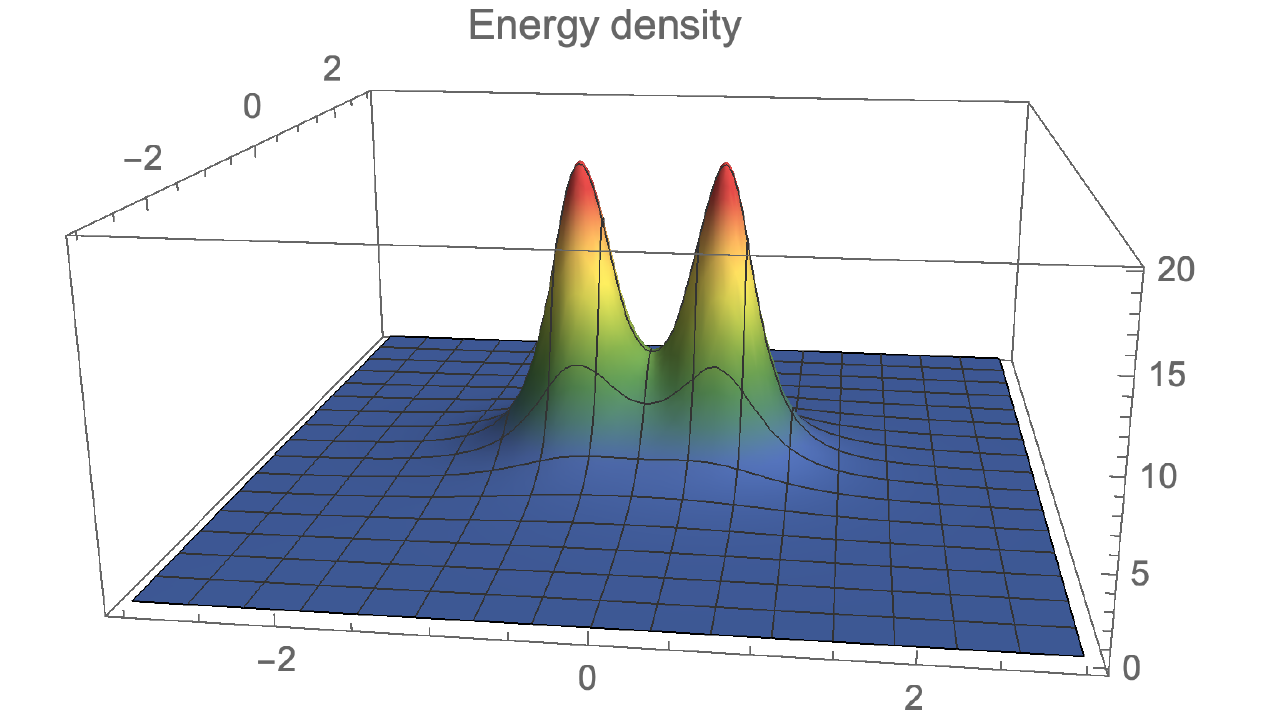}
\caption{Plots of the vortex molecule solution.
Upper-left, upper-right, and lower panels show
the Higgs field $\mathrm{Tr} |H|^2$,
$Z$-flux $Z_{xy}$, and
the energy density, respectively,
in the $xy$ plane.
The parameters are taken as the text (Eqs.~\eqref{eq:parameters1}).
The length unit is taken so that $v(=123\, \mr{GeV}) = 1$.
}
\label{Fig:molecule}
\end{figure}

\begin{figure}[tbp]
 \centering
\includegraphics[width=0.45\textwidth]{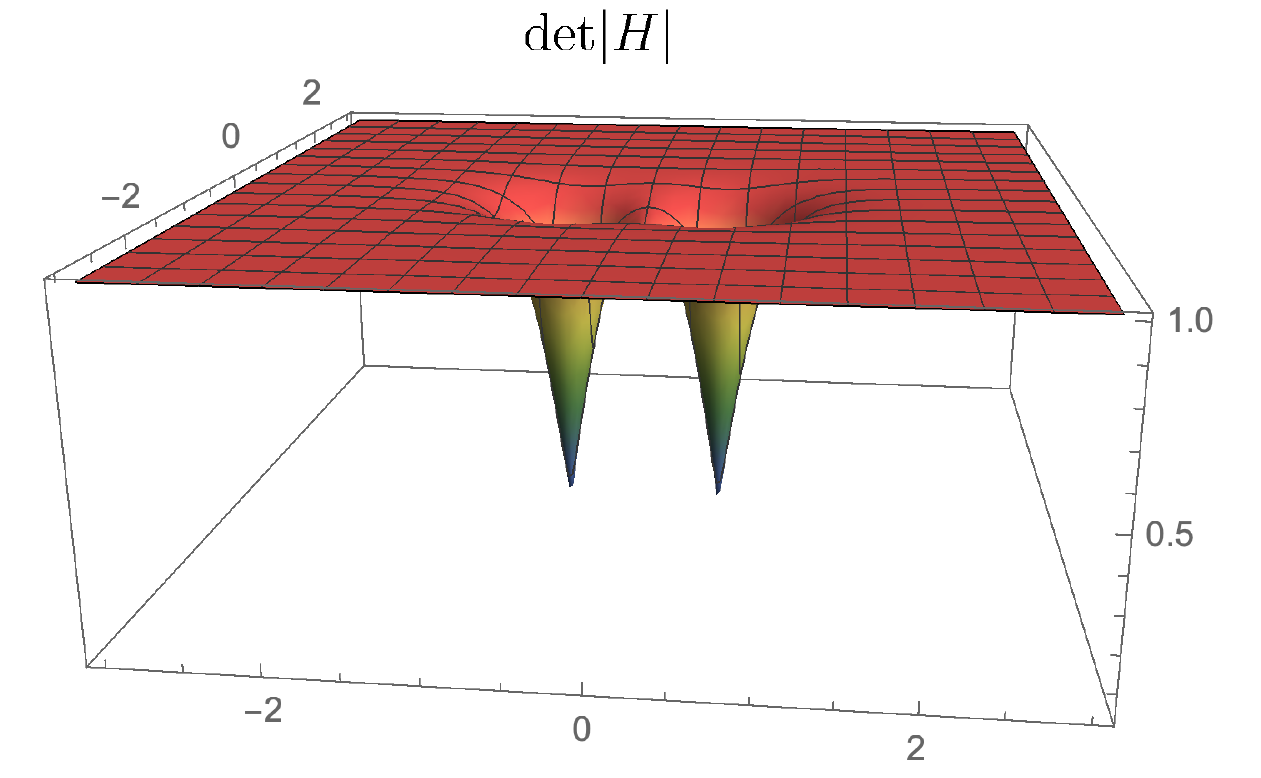} \hspace{2em}
\includegraphics[width=0.45\textwidth]{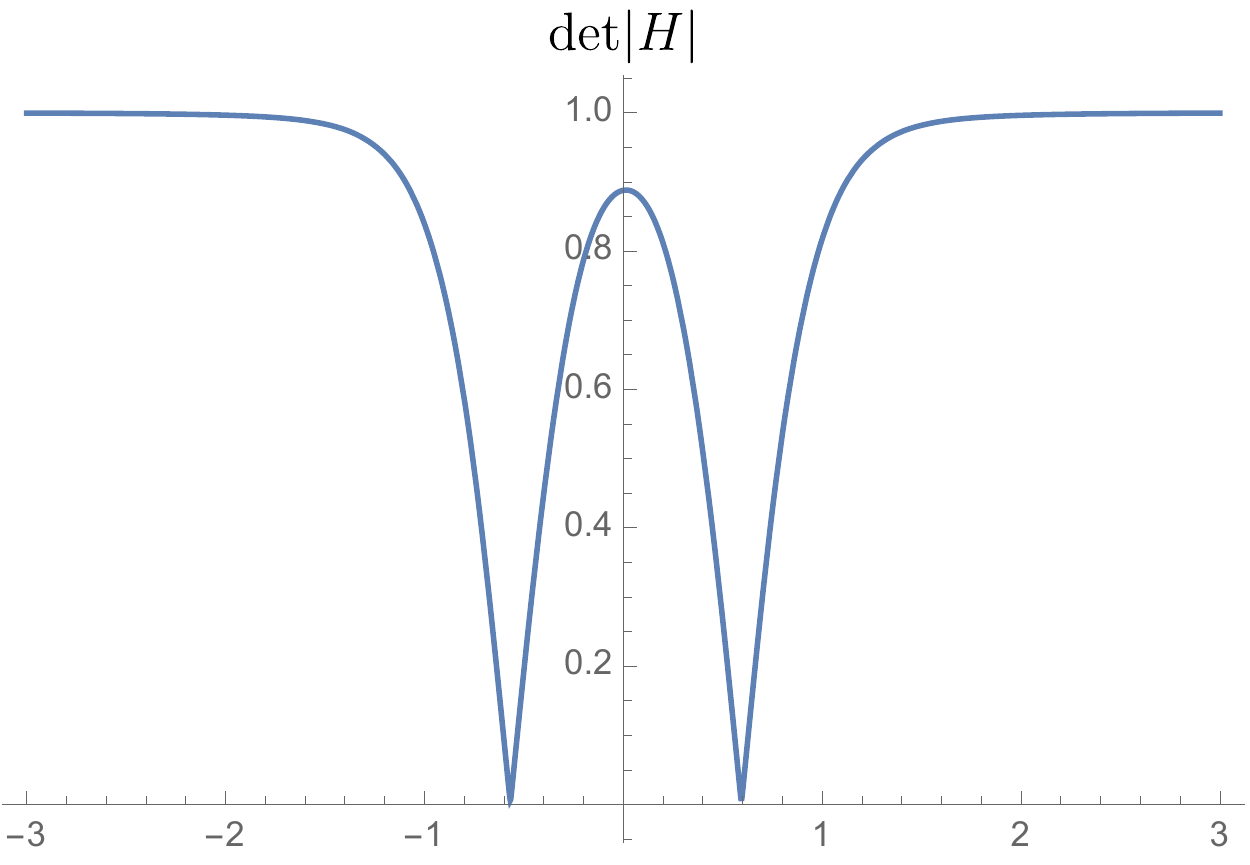} 
\caption{Plots of $\mr{det} |H|$ of the vortex molecule solution in the $U(1)_a$ symmetric case.
The right panel is the slice of the left 3D plot at $y=0$.
The determinant has two zeros at $(x,y)=(\pm 0.6 ,0)$,
indicating the positions of the two vortices.
The model parameters are the same as those in Fig.~\ref{Fig:molecule}.
}
\label{Fig:molecule2}
\end{figure}

We should note that the Higgs field does not vanish,
\textit{i.e.}, the EW symmetry is broken everywhere.
This can be seen by the upper-left panel in Fig.~\ref{Fig:molecule}, which shows the profile of $\mr{Tr}|H|^2$ of the vortex molecule.
Although it decreases around the peaks of the vortices, it does not reach to zero.
(Note that $\mr{Tr}|H|^2$ is gauge invariant.)
This explains an explicit reason of the stability of this solution,
\textit{i.e.}, it does not suffer from the condensation of the off-diagonal components in the Higgs field $H$
nor the condensation of the $W$ boson,
in contrast to the non-topological $Z$-string.

\begin{figure}[tbp]
 \centering
\begin{minipage}[]{0.45\textwidth}
\includegraphics[width=0.9\textwidth]{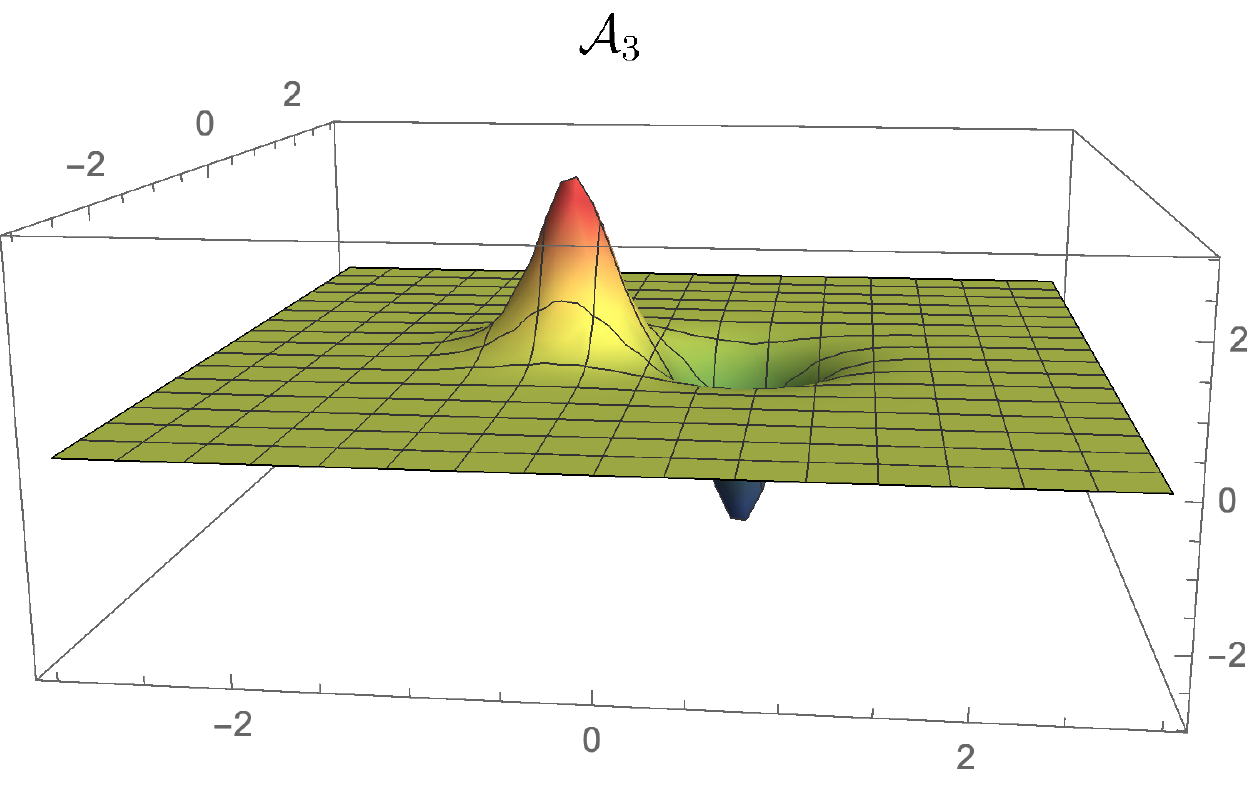}
\end{minipage}
\begin{minipage}[]{0.4\textwidth}
\vspace{2em}
\includegraphics[width=0.9\textwidth]{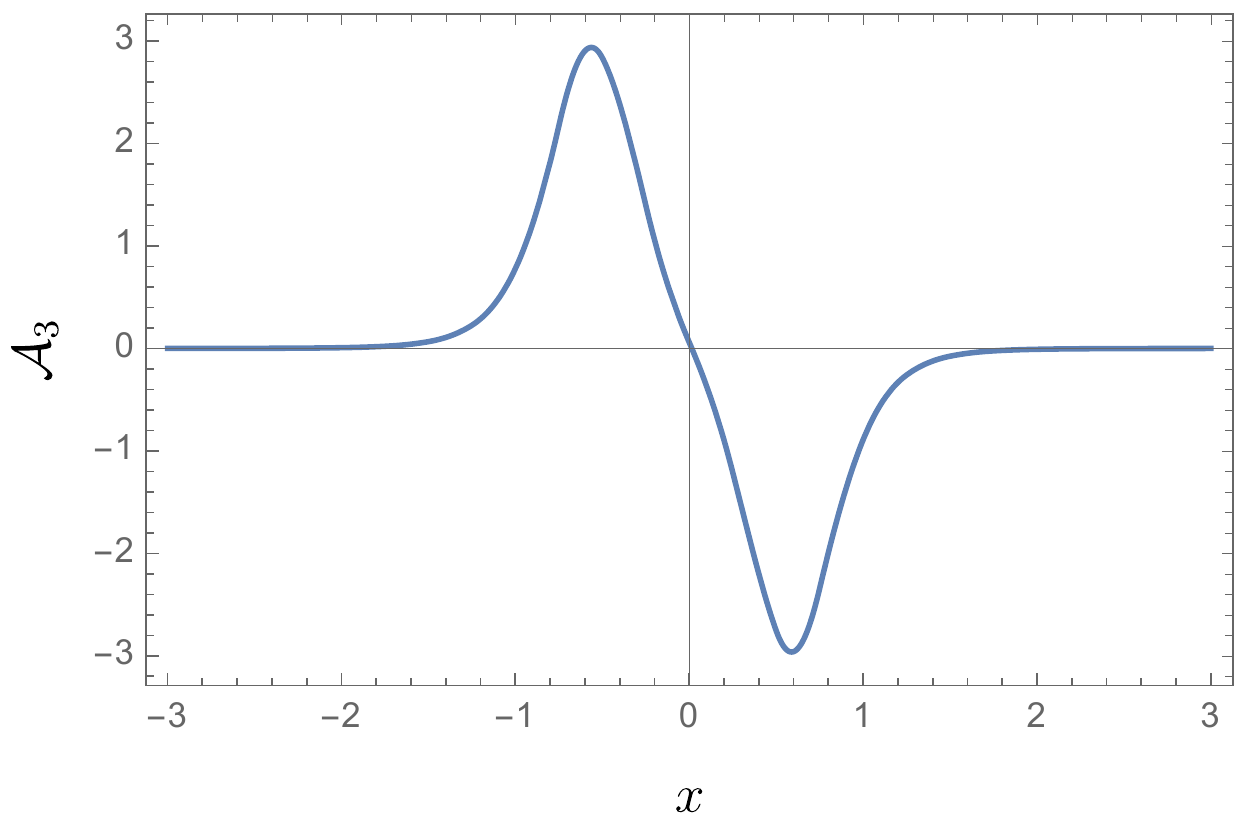}
\end{minipage}
\caption{Density of the topological charge $\mathcal{A}_3$ of $U(1)_a$ of the vortex molecule solution.
Left-panel shows 3D plot, and the right-panel shows its slice at $y=0$.
The model parameters are the same as those in Fig.~\ref{Fig:molecule}.
The length unit is taken so that $v(=123\, \mr{GeV}) = 1$.
The integrated value of $\mathcal{A}_3$ over a half-plane $x<0$ ($x>0$) is $\pm 0.4762$, respectively.
}
\label{Fig:charge}
\end{figure}

As stated above, the stability can be seen also by the viewpoint of the topological charge.
Fig.~\ref{Fig:charge} shows the topological charge density $\mathcal{A}_3$ associated with the global $U(1)_a$ symmetry,
which is defined by Eq.~\eqref{eq:topological-charge}.
Since the $U(1)_a$ symmetry is exact for the parameters Eq.~\eqref{eq:parameters1},
topological current $\mathcal{A}_3$ must be conserved.
Because the strings are well separated in the figure compared to the width of $\mathcal{A}_3$,
the pair annihilation of the topological charges is avoided, and hence this configuration is stable.

\begin{figure}[tbp]
 \centering
\includegraphics[width=0.47\textwidth]{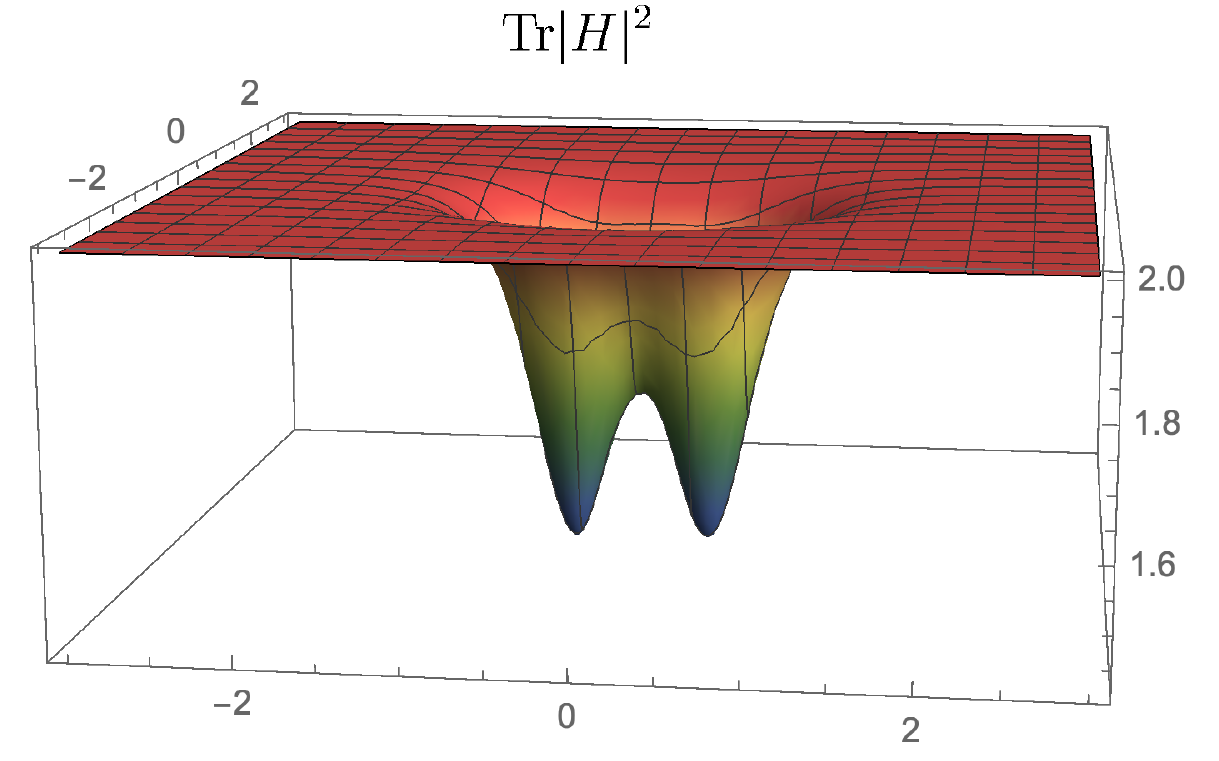} \hspace{2em}
\includegraphics[width=0.43\textwidth]{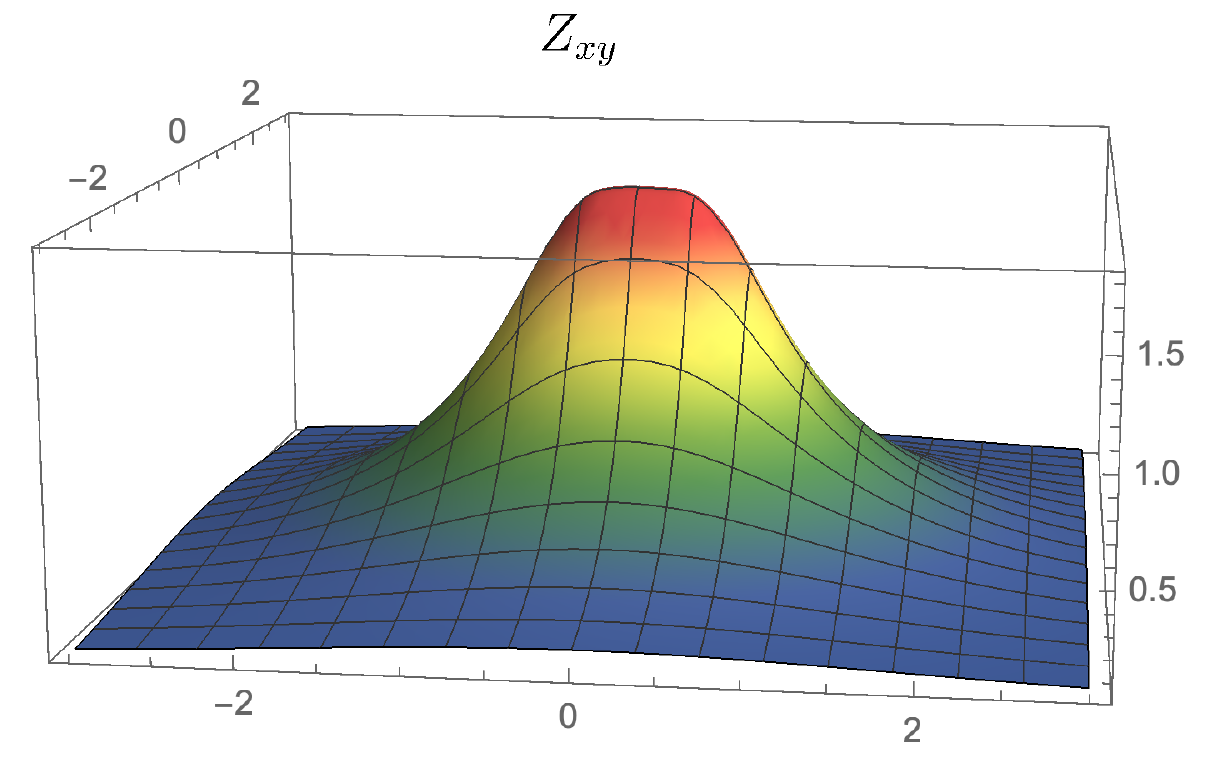} \\[5ex]
\includegraphics[width=0.45\textwidth]{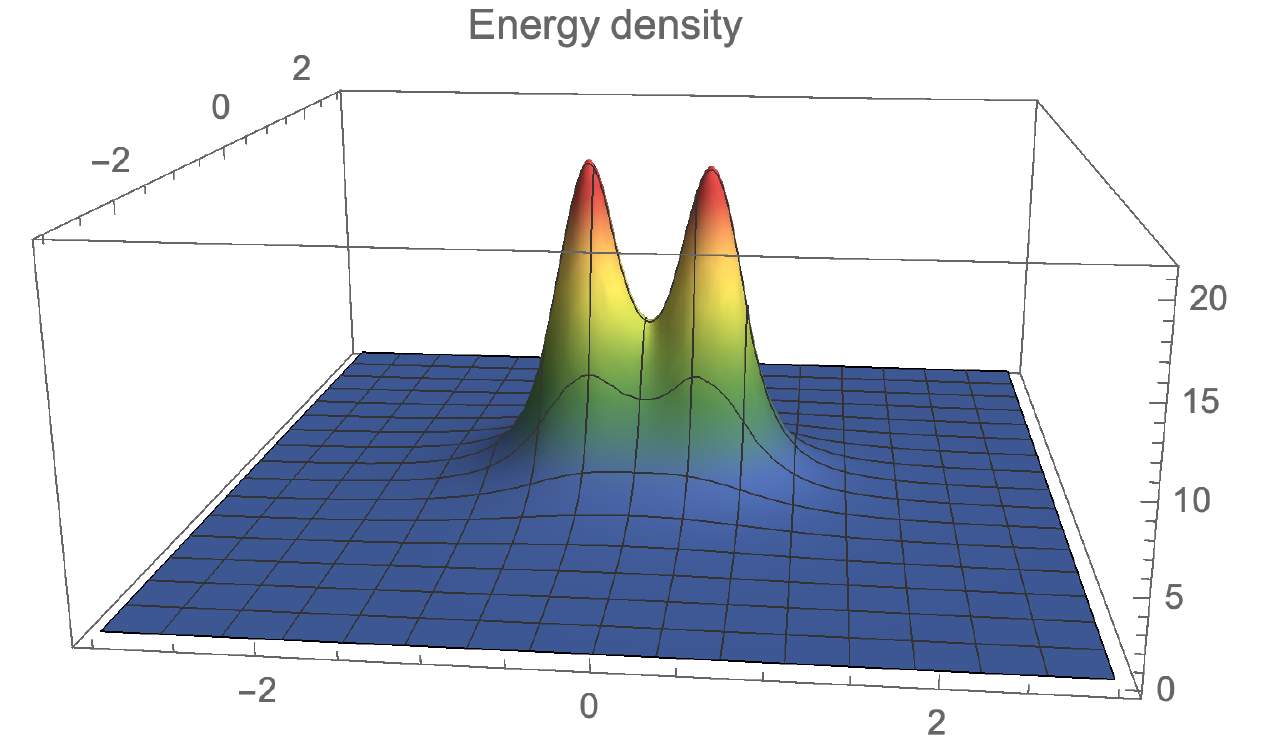}
\caption{Plots of the vortex molecule solution in the $U(1)_a$-broken case.
Upper-left, upper-right, and lower panels show
the Higgs field $\mathrm{Tr} |H|^2$,
$Z$-flux $Z_{xy}$, and
the energy density, respectively,
in the $xy$ plane.
Although the $U(1)_a$ symmetry is explicitly broken, the vortex molecule is still stable.
The parameters are taken as the text (Eq.~\eqref{eq:parameters1-wall}).
The length unit is taken so that $v(=123\, \mr{GeV}) = 1$.
}
\label{Fig:molecule-wall}
\end{figure}

As the next case, we break the $U(1)_a$ symmetry explicitly in the potential.
We do not impose the condition Eq.~\eqref{eq:cond_1}.
For simplicity, we set all parameters as real.
This condition is not essential and simply reduces the parameter space.
We take the parameters as
\begin{equation}
  m_{h^0}=1000 \, \mr{GeV} , ~  v = v_\mr{EW}/2 = 123 \, \mr{GeV} \, , \nonumber
\end{equation}
\begin{equation}
m_{H^0} = 50 \, \mr{GeV} \, , ~ m_{H^\pm}  (=m_{A^0})  =500 \, \mr{GeV}  \, ,   \nonumber
\end{equation}
\begin{equation}
  \alpha_6  =0 , ~ m_3^2 = (1 \, \mr{GeV})^2 \, , \nonumber
\label{eq:alpha6_m3}
\end{equation}
\begin{equation}
  m_Z  =91 \, \mr{GeV} , ~\sin^2 \theta_W = 0.9 \, .\label{eq:parameters1-wall}
\end{equation}
Note that $ m_{H^0} \neq 0$ (which corresponds to the pseudo NG particle of $U(1)_a$),
since the global $U(1)_a$ symmetry is explicitly broken.
From Eq.~(\ref{eq:double-SG}),
this breaking makes one domain wall attached to each fractional string \cite{Eto:2018hhg}, 
and in particular $(1,0)$ and $(0,-1)$ strings are connected by the domain wall.
Then, the long-range attractive force (a) is replaced by a confining force (a)' by the domain wall as stated in the last paragraph in Sec.~\ref{sec:interaction}. Therefore the binding force between well-separated $(1,0)$- and $(0,-1)$-strings is much stronger. 
Nevertheless,
the stable vortex molecule solution does exist
even when the domain wall is attached.

We have performed the numerical relaxation method to obtain the vortex molecule for the $U(1)_a$-broken case Eq.~\eqref{eq:parameters1-wall}.
Fig.~\ref{Fig:molecule-wall} shows plots of the obtained solution.
Again, the upper-left, upper-right, and the lower panels show
the profile of $\mr{Tr} |H|^2$, the $Z$ flux, and the energy density, respectively.
In the figure, the qualitative picture is almost the same as the case with $U(1)_a$ symmetry (Fig.~\ref{Fig:molecule}).
The only difference is that
the energy density has a wall-like structure between the two peaks corresponding to the two strings.

As stated above, the stability of the molecule is not destroyed even when $U(1)_a$ symmetry is explicitly broken.
This is because the polarization length of the molecule is too small to feel the wall tension.
In other words, the breaking effect is so small that each winding of $U(1)_a$ phase is not unwind,
which means that 
$\mathcal{A}_3$ is approximately conserved at least at classical level.
Thus the stability still holds even in this case.

\subsection{Stability region for $U(1)_a$-broken case}

We next study the parameter region for the stable molecule
when the $U(1)_a$ symmetry is explicitly broken and one domain wall is stretched between the constituent strings.
Since the molecule configuration is not axially symmetric,
performing the standard perturbation analysis becomes
very complicated, compared to those for an axially symmetric string 
done in the literature.
Therefore, instead of the small fluctuation analysis, 
we examine the stability by the relaxation method.
We start with an initial configuration with
a pair of separated $(1,0)$- and $(0,-1)$-strings.
If the molecule is stable,
it adjusts the distance between the pair and subsequently keeps the molecular shape for long time until $\partial \phi/\partial \tau$ converges
to zero within a numerical accuracy.
On the other hand, if it is unstable,
it decays into the vacuum configuration.
We repeat this procedure for various parameter
combinations to figure out parameter region
for stable molecules.

We investigate the stability for three cases with
the charged Higgs and the CP-odd neutral Higgs boson
masses $m_{H^\pm}=m_{A^0} = 110,\, 500,\ 750$
GeV. We deal with $m_{h^0}$, $\sin^2\theta_W$, and $m_{H^0}$
as free parameters, whereas we keep the custodial
symmetry condition given in Eq.~\eqref{eq:cond_2},
and $v_\mathrm{EW}$ is fixed to be 246 GeV and $m_Z=91$ GeV as before.
We also always impose $\alpha_6=0$ and $m_3^2 = (1 \, \mr{GeV})^2$.
The results are shown in Fig.~\ref{Fig:stability}.

\begin{figure}[tbp]
 \centering
\begin{minipage}[]{\textwidth}
 \centering
\includegraphics[width=0.8\textwidth]{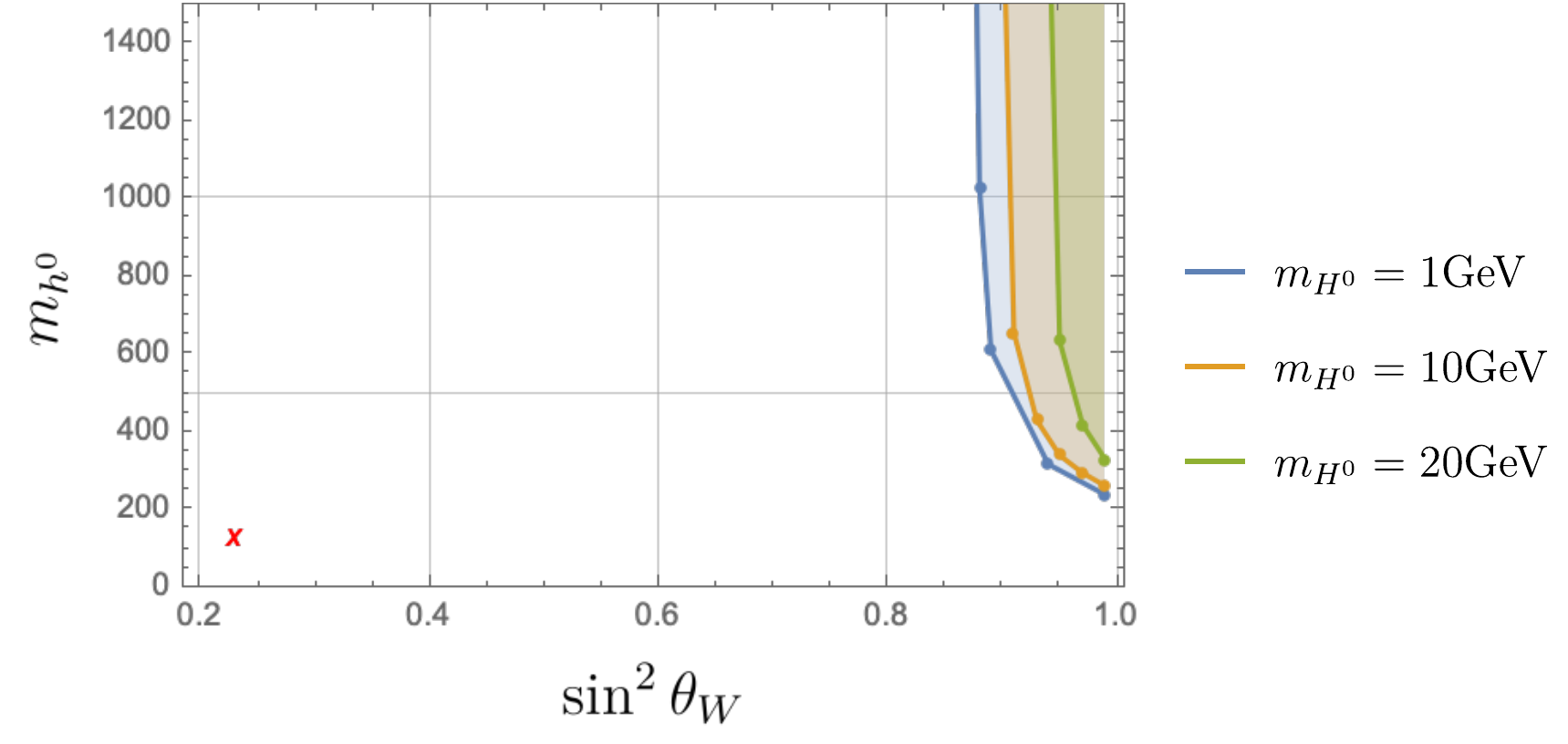}
\end{minipage}\\
\vspace{1em}
\begin{minipage}[]{\textwidth}
 \centering
\includegraphics[width=0.8\textwidth]{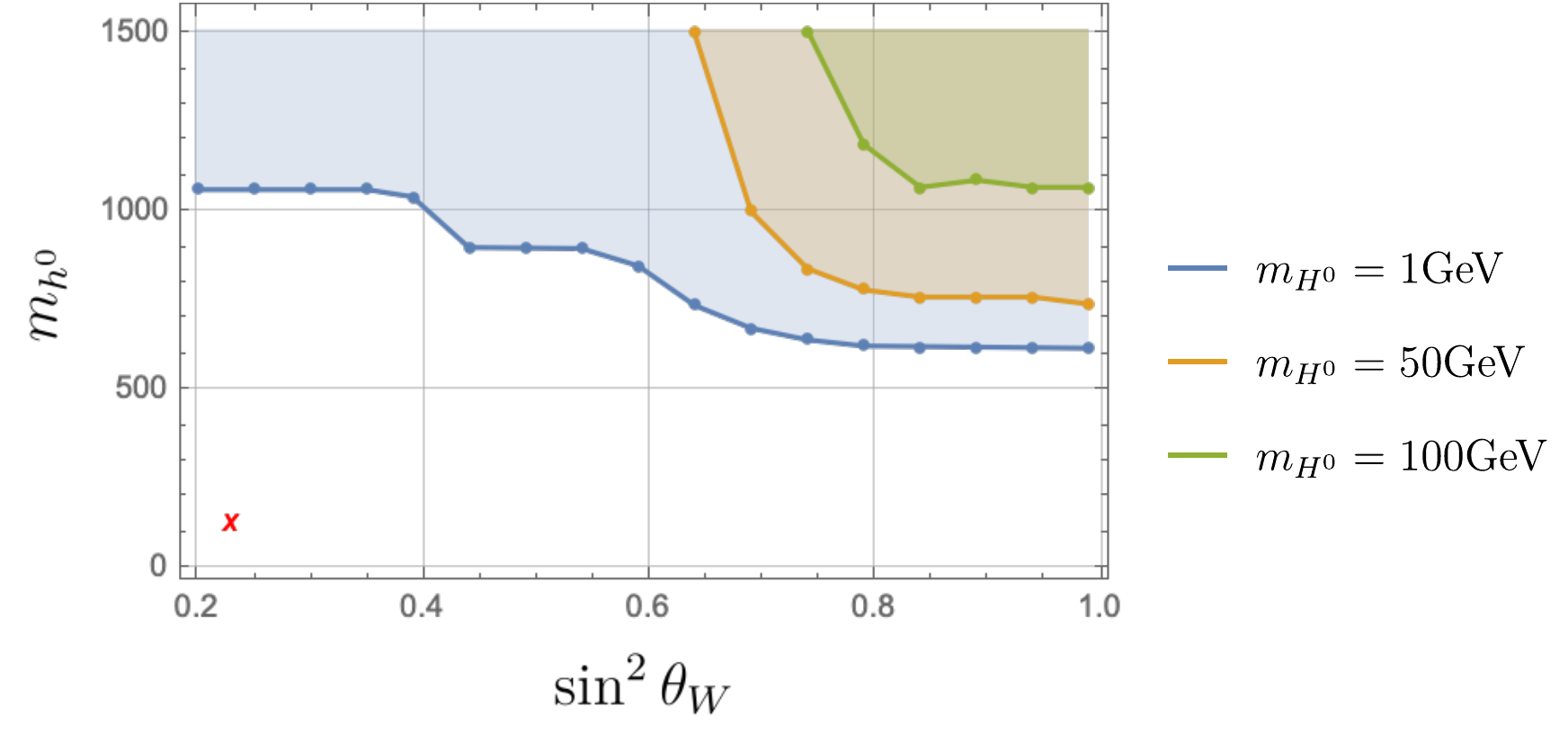}
\end{minipage}
\\
\vspace{1em}
\begin{minipage}[]{\textwidth}
 \centering
\includegraphics[width=0.8\textwidth]{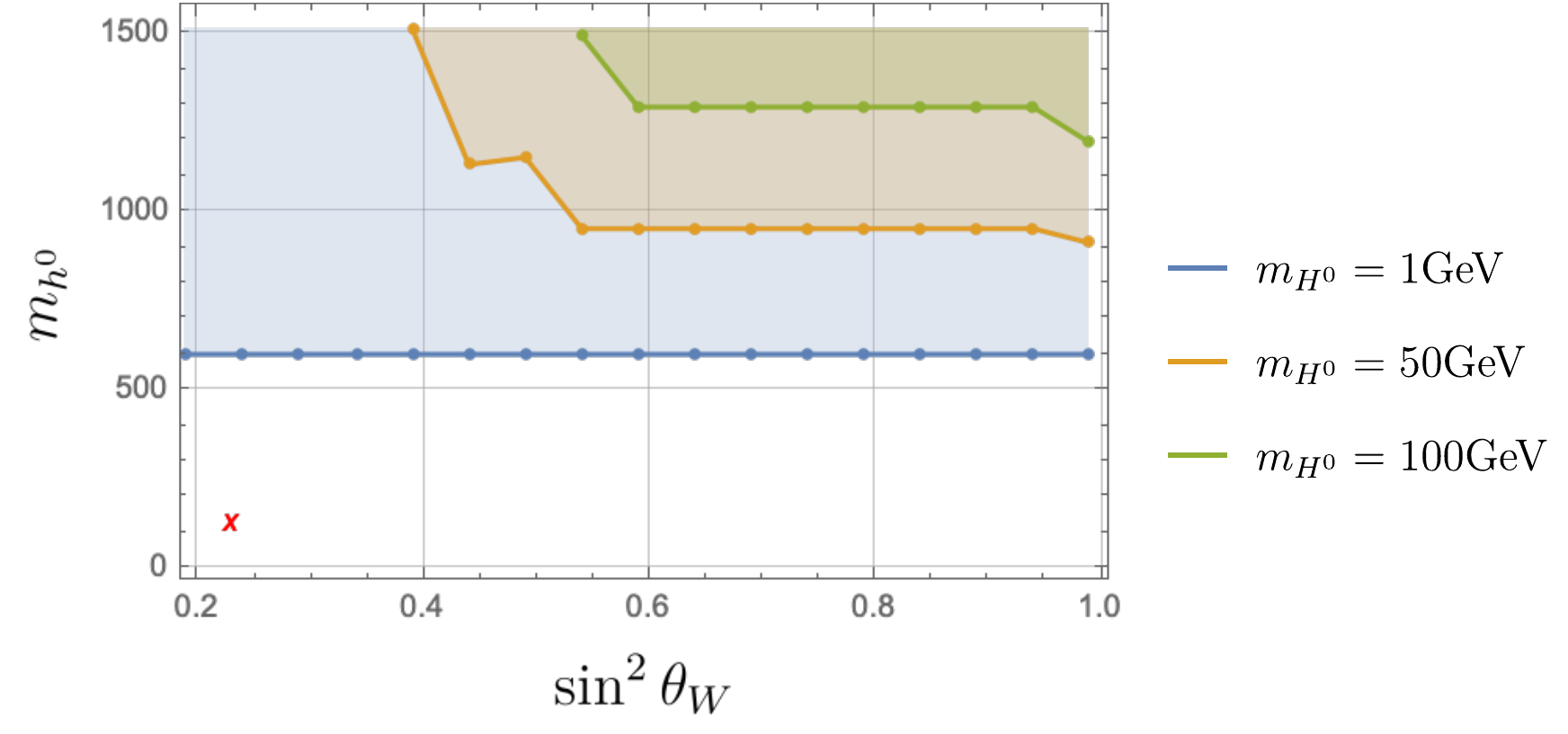}
\end{minipage}
\caption{
Stability of the molecule string. The molecule
is stable in the shaded regions. The unit of the vertical axis is GeV.
The top, middle, and bottom panels correspond to
$m_{H^\pm}=m_{A^0} = 110,\, 500,\ 750$ GeV, respectively.
The physical point $(m_{h^0},\sin^2\theta_W)=(125 \, \mr{GeV},0.23)$ is expressed by the red crosses.
}
\label{Fig:stability}
\end{figure}

For the case of $m_{H^\pm}=m_{A^0} = 110$ GeV,
we show three different curves associated
with $m_{H^0}=1,\ 10,\ 20$ GeV,
see the top panel of Fig.~\ref{Fig:stability}.
Since the mass $m_{H^0}$ of the CP-even Higgs $H^0$
is related to the domain wall tension
between the $(1,0)$- and $(0,-1)$-strings,
the smaller $m_{H^0}$ has a wider region for
the stable molecule.
We should emphasize that the usual axially symmetric
$Z$-string solution cannot be stable 
in the shaded regions of the top panel
of Fig.~\ref{Fig:stability}. The stable regions
arise due to the fact that the $Z$-string
splits into the $(1,0)$- and $(0,-1)$-strings.
However, unfortunately, the physical point with
$m_{h^0}=125$ GeV and $\sin^2\theta_W = 0.23$ is 
not included in any of the stable regions.

The similar results for $m_{H^\pm}=m_{A^0} = 500$ GeV
and 750 GeV are also shown in the middle and
bottom panels of Fig.~\ref{Fig:stability}, respectively. We plot three curves
corresponding to $m_{H^0}=1,\ 50,\ 100$ GeV in these
cases. The stable regions become larger as 
we increase $m_{H^\pm} = m_{A^0}$. 
This tendency agrees well with the qualitative condition 
in Eq.~\eqref{eq:stable-condition} given by the analytic argument.
As we described in Eq.~(\ref{eq:F_d}), the $(1,0)$-
and $(0,-1)$-strings feel a repulsive force
originated from the gradient of the profile
function $h(r)$, which must vanish at the $(1,0)$-string
center but has a large condensation $h(0)$
at the $(0,-1)$-string center. When we increase
$m_{H^\pm}=m_{A^0}$, the string core size becomes smaller,
and therefore the gradient of $h(r)$ becomes larger.
Therefore, the two constituent strings feel 
a stronger repulsive force, which prevents the molecule from shrinking.
Furthermore, due to the small string core, the topological $U(1)_a$ charge is well polarized,
resulting in the stability.

It has been known that in general
the non-topological $Z$-string is unstable in small
$\theta_W$ region since $W$-condensation
tends to occur. 
Note that the $W$-condensation can sometimes
destabilize a non-topological soliton whereas
a topologically protected soliton is always stable.
We have found a similar $\theta_W$ dependence
for the molecule but the 
instability for small $\theta_W$ 
is greatly suppressed by
increasing $m_{H^\pm}=m_{A^0}$. This is because
that each constituent string of the molecule
is locally a topological string.
Thus, the instability channel through the
$W$-condensation is suppressed for the molecule 
as long as the $(1,0)$- and $(0,-1)$-strings are well
separated from one another.

Thus, 
 unstable axially symmetric $Z$-strings 
 can become stable in the form of molecules in the 2HDM 
 in a large parameter region, 
 although we have found that the physical point
$(m_{h^0},\sin^2\theta_W)=(125\, \mr{GeV},0.23)$ always belongs
to unstable regions.


\section{Summary and discussion}
\label{sec:conclusion}
We have studied the vortex molecule, a bound state of topological $Z$-strings, in the 2HDM.
Its net topological charge is 
zero  
since it is in the same topological sector as the non-topological $Z$-string.
Nevertheless 
it can be a stable solution of the EOMs for a significantly wide parameter range.
This stability is understood by the fact that each constituent string is locally a topological string
with the $U(1)_a$ topological charge.
As long as the topological strings are separated enough by the short-range repulsive forces between them,
which are realized by the $Z$ flux and the condensation of the unwinding Higgs component in the strings,
the molecule is stabilized due to this polarization mechanism.
It is surprising that even when the global $U(1)_a$ symmetry is explicitly broken in the Higgs potential,
the vortex molecule is still stable for quite wide parameter space,
although the physical point $(m_{h^0},\sin^2\theta_W)=(125\, \mr{GeV},0.23)$ 
always belongs to unstable regions of the molecule.
While a similar idea is known for the semilocal strings \cite{Eto:2016mqc},
the present study is the first attempt to utilize the polarization to stabilize the non-topological string,
which seems to have no reason to be stable in the sense of topology.

In our study, we have imposed the $SU(2)_L\times SU(2)_R$ symmetry (broken into the custodial $SU(2)_\mathrm{C}$ symmetry) in Eq.~\eqref{eq:cond_2} 
and set all parameters are real.
(Note that the reality of the parameters does not imply the CP symmetry in our basis, the PV-II basis.)
When one relaxes these conditions, the molecule could be stable even for the physical point.
Indeed, $h^0$ and $H^0$ can be mixed at the tree level 
in the case with the complex parameters $m_3$, $\alpha_5$, and $\alpha_6$,
and are not mass eigenstates.
Their two-by-to mass matrix has two mass eigenstates, 
one of which is identified with the SM-like Higgs boson after taking the alignment limit 
($\mathrm{Im}\, \alpha_5 = - \mathrm{Im}\, \alpha_6$).
Then the stability region for the mass eigenvalues can be significantly changed in this case.
However, the number of parameters is large, and
therefore more complicated analytical/numerical analysis
is needed. Hence, we will report this
investigation 
elsewhere.

Another possibility to improve the stability of the vortex molecule is 
to extend the Higgs sector by adding a complex SM singlet scalar
and to impose the so-called global $U(1)_\mr{PQ}$ symmetry.
This model is called the DFSZ axion model \cite{Zhitnitsky:1980tq,Dine:1981rt}
and is known to have topological string configurations with a topological charge associated with the $U(1)_\mr{PQ}$ symmetry.
In addition, the strings have the electroweak gauge flux and non-zero condensation of the Higgs fields around their cores
after the electroweak symmetry is broken (called the electroweak axion string \cite{Abe:2020ure}).
Thus one can consider a similar molecule-like soliton in the DFSZ model as well.
It is expected to be more stable than the vortex molecule in 2HDM
because the $U(1)_\mr{PQ}$ symmetry is exact at the classical level.

Note that this vortex molecule is not axially symmetric,
which means that this configuration has a $U(1)$ modulus (NG mode) associated with the spontaneously broken 
$O(2)$ spatial rotation around the string.
Since this modulus is not a gauge modulus, 
the vortex molecule behaves as a superfluid string instead of a superconducting string \cite{Witten:1984eb}.
This modulus would give non-trivial dynamics for the vortex molecule.
For instance, one can twist the $U(1)$ modulus along a closed string, 
giving a loop of the vortex molecule with non-zero twist number.
This twisted $U(1)$ modulus can prevent the loop from shrinking, making it stable,
which is nothing but a kind of a vorton \cite{Davis:1988ij}.

Although we have concentrated on the stability of a single non-topological $Z$-string in the form of 
a molecule of $(1,0)$ and $(0,-1)$ topologicsl $Z$ strings, 
it will be also interesting to explore the stability of multiple molecules.
In related models admitting similar molecules mentioned below (baby Skyrme model 
\cite{Jaykka:2010bq,Kobayashi:2013aja,Kobayashi:2013wra,Akagi:2021lva}
and two-component Bose-Einstein condensates (BECs) 
\cite{Son2002,Kasamatsu2004,Eto:2012rc,Eto:2013spa,Cipriani2013,Nitta:2013eaa,Tylutki2016,Eto2018,Kobayashi:2018ezm,Eto:2019uhe}), 
multiple strings are connected to form a sheet or a polygon as (meta)stable states 
\cite{Kobayashi:2013aja,Kobayashi:2013wra}.
For sufficiently large number of molecules, the most stable configuration 
is a square lattice (the baby Skyrme model \cite{Jaykka:2010bq}) 
or both square and triangular lattices (the BEC \cite{Cipriani2013}). 
Similar situations will occur in our case of multiple $Z$-string molecules.
In such multiple configurations, the stability region may be extended compared with the single case.

As for more dynamical aspects of multiple configurations,
it is also interesting to consider the reconnection process of the vortex molecules in collisions, 
in particular for application to cosmology.
Since the molecule is a bound state of the two topological strings,
the reconnection can be considered as a four-body scattering process around the collision point.
If the reconnection probability significantly deviates from unity, 
the dynamics of the network of the vortex molecules becomes quite non-trivial 
\cite{Copeland:2006eh,Copeland:2006if,Avgoustidis:2014rqa} 
(see Ref.~\cite{Eto:2006db} for the reconnection of non-Abelian strings).
This would be useful to discuss cosmological/astrophysical consequence of the 2HDM or 
other particle physics models that have similar structures in scalar potentials.

Before closing this paper, let us discuss a possibility to apply  
our stabilization mechanism of the topological polarization 
to other non-topological strings and solitons.
As mentioned in introduction, $Z$-strings reduce to semilocal strings for $\theta_W = \pi/2$ \cite{Vachaspati:1991dz,Achucarro:1999it}, 
and stabilizing the semilocal strings by polarization was discussed in Ref.~\cite{Eto:2016mqc}.
However, this was not the same with ours in the sense that 
the stabilizing term makes strings topological.
In addition to this example, 
semilocal strings are further reduced to lump-strings 
in the strong $U(1)$ gauge coupling limit, in which the model reduces to the $O(3)$ sigma model.
With the same potential with the above mentioned semilocal theory, 
a single lumps-string is split into two fractional lump-strings again forming a molecule
\cite{Schroers:1995he,Schroers:1996zy,Nitta:2011um}.
In condensed matter physics, this is actually known in ferromagnets 
in which the potential is called an easy-plane potential.
The same happens for baby-Skyrmions in a baby Skyrme model, 
the $O(3)$ sigma model with four-derivative term, 
with an easy-plane potential term \cite{Jaykka:2010bq,Kobayashi:2013aja,Kobayashi:2013wra,Akagi:2021lva}.
In these cases, lumps and baby-Skyrmions are both topological solitons 
supported by a topological charge defined by 
the second homotopy class $\pi_2$ (instead of $\pi_1$ for vortices).
Thus, in this sense, these are different from our case of 
the stabilization mechanism of splitting non-topological solitons.
Application of our stabilization mechanism of the topological polarization to 
other non-topological solitons such as Q-balls is a future problem

Let us make further comments 
on fractional vortex molecules in other cases.
Actually, a vortex molecule consisting of fractional vortices 
connected by a domain wall can be often seen in 
various multicomponent condensed matter systems such as 
two-gap (or two-component) superconductors \cite{Babaev:2001hv,
Goryo2007,Tanaka2007,Crisan2007,Guikema2008,Garaud:2011zk,Garaud:2012pn,Tanaka2017,Tanaka2018,Chatterjee:2019jez}, 
coherently coupled two-component BECs \cite{Son2002,Kasamatsu2004,Eto:2012rc,Eto:2013spa,Cipriani2013,Nitta:2013eaa,Tylutki2016,Eto2018,Kobayashi:2018ezm,Eto:2019uhe}, 
and a color superconductor of dense QCD \cite{Eto:2021nle}.
However, a crucial difference is that 
in these systems the total configurations posses nonzero topological charges. 
Thus, even when the attraction between constituent fractional vortices dominates 
leading molecules to collapse, there remain singly quantized topological vortices.
On the contrary, in our case, 
the total topological charge vanishes and the collapse of molecules implies a decay ending up with the vacuum. 
Therefore, the topological polarization to stabilize non-topological strings that we have found 
in this paper is novel and more challenging than these cases.

Finally, it will be also interesting to investigate vortex molecules in multi-Higgs doublet models.
For an $N$-Higgs doublet model, the molecule would be composed of $N$ fractional strings with $1/N$ fractional $Z$ fluxes.

\section*{Acknowledgement}
Y.H. would like to thank Masashi Aiko for useful discussions.
%
This work is supported in part by JSPS Grant-in-Aid for Scientific Research 
(KAKENHI Grant
 No.~JP19K03839 (M.~E.),
 No.~JP21J01117 (Y.~H.),
 No.~JP18H01217 (M.~N.)), 
and also by MEXT KAKENHI Grant-in-Aid for Scientific Research on Innovative Areas
``Discrete Geometric Analysis for Materials Design'' No.~JP17H06462 (M.~E.)
 from the MEXT of Japan.

\appendix

\section{Difference of conventions from previous works}
\label{app:basis}

In Refs.~\cite{Eto:2019hhf,Eto:2020hjb,Eto:2020opf}, 
the Nambu monopole in 2HDM is studied in a convention that is different from the present paper.
There, a global $SU(2)$ symmetry and its discrete $\mathbb{Z}_2$ subgroup symmetry for the Higgs potential are introduced.
The former is called the custodial $SU(2)_\mr{C}$ symmetry in Refs.~\cite{Eto:2019hhf,Eto:2020hjb,Eto:2020opf}
and is completely the same as Eq.~\eqref{eq:custodial2}, 
\textit{i.e.}, the custodial symmetry of the case II studied by Pomarol and Vega~\cite{Pomarol:1993mu}.
The latter $\mathbb{Z}_2$ symmetry is called the $(\mathbb{Z}_2)_\mr{C}$ symmetry in Refs.~\cite{Eto:2019hhf,Eto:2020hjb,Eto:2020opf}.
However, it is equivalent to Eq.~\eqref{eq:CP2} in this paper (up to a constant gauge transformation),
and thus should be regarded as the CP symmetry in the PV-II basis.
Instead of that, the CP symmetry in Refs.~\cite{Eto:2019hhf,Eto:2020hjb,Eto:2020opf} is defined as the charge conjugation of the Higgs fields, which is not a $\mathbb{Z}_2$ subgroup of the custodial symmetry.
Consequently, some terminologies on the CP symmetry are different from this paper.

It should be useful to summarize their correspondence as the following table:

\begin{table}[h]
  \centering
  \begin{tabular}{ccc}
   \hline
   terminologies in Refs.~\cite{Eto:2019hhf,Eto:2020hjb,Eto:2020opf} && terminologies in this paper \\
   \hline \hline
   $(\mathbb{Z}_2)_\mr{C}$ symmetry & =& CP symmetry in the PV-II basis (Eq.~\eqref{eq:CP2}) \\
   \hline
   CP symmetry & =& complex conjugate in the PV-II basis  \\
   \hline
   CP-odd Higgs boson $A$  & =& CP-even Higgs boson $H^0$ \\
   \hline
   CP-even Higgs boson $H$  & =& CP-odd Higgs boson $A^0$  \\
   \hline
  \end{tabular}
\end{table}

Note that the $U(1)_a$ symmetry in Refs.~\cite{Eto:2019hhf,Eto:2020hjb,Eto:2020opf} is completely the same as that in this paper, Eq.~\eqref{eq:U1a}.
It is shown in Ref.~\cite{Eto:2019hhf} that 
the Nambu monopole in 2HDM can be topologically stable when the Higgs potential has $U(1)_a$ and $(\mathbb{Z}_2)_\mr{C}$ symmetries,
latter of which is equivalent to the CP symmetry as stated above.
Therefore, the stability of the Nambu monopole does not require any ad-hoc $\mathbb{Z}_2$ symmetry.
What is necessary to ensure its stability is only the $U(1)_a$ and CP symmetries in the present paper.

\section{Stability of $Z$-string in SM}
\label{app:stability-Zstring}

We give a brief review on the stability of the $Z$-string in the SM
based on Refs.~\cite{Vachaspati:1992fi,Vachaspati:1992jk,Barriola:1993fy,Perkins:1993qz}. 
The Lagrangian of the EW sector of the SM is given by
\begin{equation}
 \mathcal{L}=  - \frac{1}{4}Y_{\mu\nu}Y^{\mu\nu} -\frac{1}{2}\mr{Tr} ~ W_{\mu\nu}W^{\mu\nu}  + |D_\mu \Phi|^2 - \lambda \left(|\Phi|^2 - \frac{v_\mr{EW}^2}{2}\right)^2
\end{equation}
where $\Phi$ is the Higgs doublet with the hypercharge $+1$,
and $W_{\mu\nu}$, $Y_{\mu\nu}$ and the covariant derivative are the same as those in Sec.~\ref{sec:model}.
The $Z$-string in the SM is described by the following ansatz:
\begin{equation}
 \Phi = \frac{v_\mr{EW}}{\sqrt{2}} 
\begin{pmatrix}
 0 \\ f_\mr{SM} (r) e^{i\varphi}
\end{pmatrix} \, ,\label{eq:Zstring-Higgs}
\end{equation}
\begin{equation}
 Z_i = -\frac{2 \cos \theta_W}{g} \frac{\epsilon_{3ij}x^j}{r^2} (1-w_\mr{SM}(r)) \, \label{eq:Zstring-gauge}
\end{equation}
in the polar coordinates $(r,\varphi)$.
The profile functions satisfy the following boundary condition
\begin{equation}
 f_\mr{SM}(0) = 0, ~ w_\mr{SM}(0)=1, ~ f_\mr{SM}(\infty) =1, ~ w_\mr{SM}(\infty) = 0 \, .
\end{equation}
This configuration has a singly quantized $Z$ flux
\begin{equation}
 \Phi_Z = \frac{4\pi \cos \theta_W}{g}.
  \label{eq:single}
\end{equation}

The $Z$-string in Eqs.~\eqref{eq:Zstring-Higgs} and \eqref{eq:Zstring-gauge} is a solution of the EOMs
when the functions $f_\mr{SM}$ and $w_\mr{SM}$ are taken appropriately
because this is nothing but the embedding of the ANO string solution into the SM.
The stability is, however, not ensured in general
because of the triviality of the vacuum topology of the SM.
In fact, it is known that the $Z$-string has two unstable directions around it.

To illustrate this, let us first consider a perturbation for the Higgs doublet around the $Z$ string configuration,
\begin{equation}
 \Phi(\xi) = \frac{v_\mr{EW}}{\sqrt{2}} 
\begin{pmatrix}
 0 \\ f_\mr{SM} (r) e^{i\varphi}
\end{pmatrix}  + \frac{v_\mr{EW}}{\sqrt{2}} 
\begin{pmatrix}
\xi \\ 0
\end{pmatrix}  \, ,
\end{equation}
with a sufficiently small constant $\xi$.
Substituting this into the Higgs potential,
we obtain
\begin{align}
 V &= \lambda \frac{v_\mr{EW}^4}{4} \left(f_\mr{SM}(r)^2 + \xi^2 - 1\right)^2  \nonumber \\
 & =\lambda \frac{v_\mr{EW}^4}{4} \left(f_\mr{SM}(r)^2 - 1\right)^2  + \lambda \frac{v_\mr{EW}^4}{2} \left(f_\mr{SM}(r)^2 - 1\right) \xi^2 + \mathcal{O}(\xi^4) \, .\label{eq:instability_Higgs}
\end{align}
It is clear that, at the center of the $Z$-string, $\xi$ feels a tachyonic mass since $f_\mr{SM}$ vanishes there,
which implies the existence of instability in the direction of $\xi$.
Eventually, this instability results in forming a condensation of the component in the direction of $\xi$
and the $Z$ string decays as the $Z$ flux is infinitely diluted.

There is another instability for the $Z$-string.
Let us focus on the $W^\pm$ boson,
which has an interaction vertex as
\begin{equation}
\mathcal{L}_W \supset i g \cos\theta_W Z_{\mu\nu} W^{-,\mu}W^{+,\nu} + \mr{h.c.}\, 
\end{equation}
where $Z_{\mu\nu}$ is the field strength of the $Z$ boson.
In the $Z$-string background, Eq.~\eqref{eq:Zstring-gauge},
$Z_{\mu\nu}$ is given as 
\begin{equation}
 Z_{12} = \frac{2 \cos \theta_W}{g} \frac{w_\mr{SM}'(r)}{r} \, ,
\end{equation}
and together with the quadratic mass term 
\begin{equation}
 \frac{g^2}{2} |\Phi|^2 \, W_\mu^- W^{+,\mu},
\end{equation}
we have the following two-by-two mass matrix for the $W$ boson:
\begin{equation}
 \mathcal{L}_\text{mass} =  - 
\begin{pmatrix}
W_1^-  & W_2^-
\end{pmatrix}
\begin{pmatrix}
 \frac{g^2}{2} |\Phi|^2  & - ig \cos \theta_W Z_{12} \\   ig \cos \theta_W Z_{12} &  \frac{g^2}{2} |\Phi|^2
\end{pmatrix}
\begin{pmatrix}
W_1^+  & W_2^+
\end{pmatrix} \, .
\end{equation}
It is convenient to take two linear combinations,
\begin{equation}
 W^{\uparrow} \equiv \frac{1}{\sqrt{2}}(W_1^+ + i W_2 ^+) , \quad
 W^{\downarrow} \equiv \frac{1}{\sqrt{2}}(W_1^+ - i W_2 ^+) \, ,
\end{equation}
and then the mass term is rewritten as
\begin{align}
 \mathcal{L}_\text{mass} =  &
- \left(\frac{g^2}{2} |\Phi|^2 + g \cos \theta_W Z_{12} \right) |W^\uparrow|^2
- \left(\frac{g^2}{2} |\Phi|^2 - g \cos \theta_W Z_{12} \right) |W^\downarrow|^2 \nonumber\\
= &
- \left(m_W^2 f_\mr{SM}(r)^2 + 2\cos^2 \theta_W \frac{w_\mr{SM}'(r)}{r}\right) |W^\uparrow|^2 \nonumber \\
& \hspace{5em} 
- \left(m_W^2 f_\mr{SM}(r)^2 - 2\cos^2 \theta_W \frac{w_\mr{SM}'(r)}{r} \right) |W^\downarrow|^2
\end{align}
where we have substituted the ansatz Eq.~\eqref{eq:Zstring-Higgs} 
and used $m_W ^2 = g^2 v_\mr{EW}^2/4$.
It is clear that, at the center of the $Z$-string
in which $f_\mr{SM}(r)$ vanishes,
the polarization mode $W^\uparrow$ feels the tachyonic mass (note that $w_\mr{SM}'(r)$ is negative),
which means that the $W$ boson tends to condensate around the $Z$-string (see also Refs.~\cite{Ambjorn:1989sz,Ambjorn:1989bd}).
This is the second instability
and the $Z$-string can decay by this process.

Although the above argument is illustrative,
it only states that there are two dangerous directions potentially leading to the instability.
To clarify the domain of the stability in the parameter space,
one has to solve the Schr\"odinger-like equation with respect to linearized perturbations around the $Z$-string configuration
and obtain its eigenvalues.
Such studies are found in Refs.~\cite{Vachaspati:1992jk,Earnshaw:1993yu,James:1992wb,Goodband:1995he}.
As a result, it is known that the $Z$-string can be a stable solution only when the Weinberg angle $\theta_W $ is close to $\pi/2$ and
the Higgs boson mass is smaller than that of the $Z$ boson.
Indeed, for $\theta_W =\pi/2$,
the $Z$-string reduces to a semilocal string \cite{Vachaspati:1991dz,Achucarro:1999it}, 
the $SU(2)_W$ gauge fields are decoupled, and hence the $W$ condensation does not occur.
In addition, when the Higgs boson is lighter than the $Z$ boson,
the Higgs potential is insignificant compared to its kinetic term
and the tachyonic mass found in Eq.~\eqref{eq:instability_Higgs} is not relevant.


\section{Derivation of interaction between topological $Z$-strings}
\label{app:interaction}
We here give details of the computations in Sec.~\ref{sec:interaction}.
The $(1,0)$- and $(0,-1)$-strings are put as Fig.~\ref{fig:two-strings}.

Firstly let us derive Eq.~\eqref{eq:deltaT-a}.
To do so, we calculate the kinetic energy of the Higgs field,
\begin{equation}
 K_H= \mr{Tr}|D_i H|^2 .
\end{equation}
Using Eq.~\eqref{170642_15Jun21}, the covariant derivative 
$D_i H$ can be calculated as
\begin{align}
 D_i H &= v  \,
\mr{diag.}
\begin{pmatrix}
\left[\vec{\nabla}(f(r_1) h(r_2))  + i f(r_1) h(r_2) \left(\vec{\nabla}\theta_1  - \frac{g_Z }{2}\vec{Z}\right)\right] e^{i \theta_1} \\
\left[\vec{\nabla}(f(r_2) h(r_1))  - i f(r_2) h(r_1) \left(\vec{\nabla}\theta_2  - \frac{g_Z }{2} \vec{Z}\right) \right] e^{-i \theta_2}
\end{pmatrix}
\end{align}
and we obtain 
\begin{align}
\mr{Tr}|D_i H|^2 & = v^2 \left[\vec{\nabla}(f(r_1) h(r_2))  \right]^2 
+ v^2 (f(r_1) h(r_2))^2 \left(\vec{\nabla}\theta_1  - \frac{g_Z }{2}\vec{Z}\right)^2 \n \\
& + v^2 \left[\vec{\nabla}(f(r_2) h(r_1))  \right]^2 
+ v^2 (f(r_2) h(r_1))^2 \left(\vec{\nabla}\theta_2  - \frac{g_Z }{2}\vec{Z}\right)^2 .
\end{align}
We are interested in the second and the fourth terms,
which contribute to the long-range force (a).
On the other hand, for the single $(1,0)$ and $(0,-1)$-strings,
we have
\begin{align}
\mr{Tr}|D_i H^{(1,0)}|^2 & = v^2 \left[\vec{\nabla}(f(r_1))  \right]^2 
+ v^2 f(r_1)^2 \left(\vec{\nabla}\theta_1  - \frac{g_Z }{2}\vec{Z}^{(1,0)}\right)^2 \n \\
& + v^2 \left[\vec{\nabla}h(r_1)  \right]^2 
+ v^2 h(r_1)^2 \left(\frac{g_Z }{2}\vec{Z}^{(1,0)}\right)^2 .
\end{align}
and
\begin{align}
\mr{Tr}|D_i H^{(0,-1)}|^2 & = v^2 \left[\vec{\nabla}h(r_2)  \right]^2 
+ v^2 h(r_2)^2 \left(\frac{g_Z }{2}\vec{Z}^{(0,-1)}\right)^2 \n \\
& + v^2 \left[\vec{\nabla}f(r_2)   \right]^2 
+ v^2 f(r_2)^2 \left(\vec{\nabla}\theta_2  - \frac{g_Z }{2}\vec{Z}^{(0,-1)}\right)^2 .
\end{align}
Recalling Eq.~\eqref{201102_22Jun21}, we have
\begin{align}
&  \left(\vec{\nabla}\theta_1  - \frac{g_Z }{2}\vec{Z}\right)^2 \n \\
 & = \f{1}{4 r_1^2} + \f{1}{4 r_2^2} -\f{1}{2 r_1 r_2}\cos (\theta_1-\theta_2) 
+ \mathcal{O}(\delta w(r_1), \delta w(r_2) )\, .\label{eq:asym-winding}
\end{align}
and 
\begin{equation}
 \left(\vec{\nabla}\theta_1  - \frac{g_Z }{2}\vec{Z}^{(1,0)}\right)^2 = \f{1}{4 r_1^2} + \mathcal{O}(\delta w(r_1) ) \, ,
\end{equation}
\begin{equation}
 \left(\vec{\nabla}\theta_2  - \frac{g_Z }{2}\vec{Z}^{(0,-1)}\right)^2 = \f{1}{4 r_2^2} + \mathcal{O}(\delta w(r_2) ) \,,
\end{equation}
and $\left(\vec{\nabla}\theta_2  - \frac{g_Z }{2}\vec{Z}\right)^2$ has the same form as Eq.~\eqref{eq:asym-winding}.
We have used $\vec e_{\theta_1}\cdot \vec e _{\theta_2}=\cos (\theta_1-\theta_2)$.
On the other hand, 
using the asymptotic expressions of $f$ and $h$,
Eqs.~\eqref{eq:asymp1_F}-\eqref{eq:asymp2_G} and Eq.~\eqref{eq:new-profile-function},
we have
\begin{align}
 (f(r_1) h(r_2))^2 & 
= \left[\left(1+\f{1}{2} \delta F(r_1) + \f{1}{2}\delta G(r_1)\right)\left(1+\f{1}{2} \delta F(r_2) - \f{1}{2}\delta G(r_2)\right)\right]^2 \n \\
& = \left[1 + \f{1}{2}\left( \delta F(r_1) + \delta F(r_2) + \delta G(r_1) - \delta G(r_2)\right) + \mathcal{O}(\delta F^2, \delta G^2)\right]^2 \n \\
& = 1 + \delta F(r_1) + \delta F(r_2) + \delta G(r_1) - \delta G(r_2) + \mathcal{O}(\delta F^2, \delta G^2)\label{eq:asymp-fh} 
\end{align}
and thus obtain
\begin{align}
 \delta T \big |_\text{(a)} 
& \simeq \int d^2 x \, v^2
\Big[
 (f(r_1) h(r_2))^2 \left(\vec{\nabla}\theta_1  - \frac{g_Z }{2}\vec{Z}\right)^2 
+ (f(r_2) h(r_1))^2 \left(\vec{\nabla}\theta_2  - \frac{g_Z }{2}\vec{Z}\right)^2 \n 
\nonumber
\\
& \h{3em} - f(r_1)^2  \left(\vec{\nabla}\theta_1  - \frac{g_Z }{2}\vec{Z}^{(1,0)}\right)^2 
- h(r_1)^2 \left(\frac{g_Z }{2}\vec{Z}^{(1,0)}\right)^2 \nonumber\\
& \h{3em} - f(r_2)^2 \left(\vec{\nabla}\theta_2  - \frac{g_Z }{2}\vec{Z}^{(0,-1)}\right)^2 
- h(r_2)^2 \left(\frac{g_Z }{2}\vec{Z}^{(0,-1)}\right)^2 
\Big] 
\nonumber\\
&= \int d^2 x \, v^2
\Big[
\f{1}{2 r_1^2} + \f{1}{2 r_2^2} -\f{1}{r_1 r_2}\cos (\theta_1-\theta_2) 
-\f{1}{2 r_1^2} -\f{1}{2 r_2^2} + \mathcal{O}(\delta F, \delta G, \delta w)
\Big]
\nonumber\\
&= \int d^2 x \, v^2
\Big[
 -\f{1}{r_1 r_2}\cos (\theta_1-\theta_2) + \mathcal{O}(\delta F, \delta G, \delta w)
\Big] \, ,
\end{align}
which coincides with Eq.~\eqref{eq:deltaT-a}.

Next, let us derive Eq.~\eqref{eq:deltaT-b}.
To do so, it is sufficient to concentrate on terms involving $\delta F$ linearly.
(Terms involving its derivative $\delta F '$ are sub-leading with respect to $R$.)
Such terms come both from the kinetic term $K_H$ and the potential energy $V$.
Using Eqs.~\eqref{eq:asym-winding}-\eqref{eq:asymp-fh},
we have
\begin{align}
 K_H  \big|_\text{linear $\delta F$} 
 = & v^2 \left(\delta F(r_1) + \delta F(r_2)\right)\left[\f{1}{2 r_1^2} + \f{1}{2 r_2^2} -\f{1}{r_1 r_2}\cos (\theta_1-\theta_2) \right] \n \\
& + \mathcal{O}(\delta F^2, \delta F' , \delta G, \delta w) 
\end{align}
Similarly, the kinetic terms of the single $(1,0)$ and $(0,-1)$ strings are given as
\begin{align}
 K_H^{(1,0)}\big|_\text{linear $\delta F$} 
& = \mr{Tr}|D_i H^{(1,0)}|^2 \nonumber\\
&= v^2 \delta F(r_1) \f{1}{2 r_1^2} + \mathcal{O}(\delta F^2,\delta F', \delta G,  \delta w),
\end{align}
\begin{align}
 K_H^{(0,-1)}\big|_\text{linear $\delta F$} 
& = \mr{Tr}|D_i H^{(0,-1)}|^2 \nonumber\\
&= v^2 \delta F(r_2) \f{1}{2 r_2^2} + \mathcal{O}(\delta F^2,\delta F', \delta G,  \delta w) \, ,
\end{align}
respectively. 
On the other hand,
the calculation of the potential energy $V$ is straightforward 
and similar to that for two half-quantized vortices in 2-component BEC \cite{Eto:2011wp}.
Then, the difference of the potential energy from the two single vortices can be calculated as
\begin{align}
 \delta V &\equiv V - V^{(1,0)} - V^{(0,-1)}\nonumber \\ 
& = 2 m_1^2 v^2 \delta F(r_1) \delta F(r_2) 
 %
\end{align}
Thus, we obtain
\begin{align}
  \delta T \big |_\text{(b)} 
& \simeq \int d^2 x \, v^2
\left[
\f{\delta F(r_2)}{2 r_1^2} + \f{\delta F(r_1)}{2 r_2^2} - \left(\delta F(r_1) + \delta F(r_2)\right) \f{1}{r_1 r_2}\cos (\theta_1-\theta_2) \right. \nonumber\\
& \left. \h{8em} + 2 m_1^2 v^2 \delta F(r_1) \delta F(r_2)  
+ \mathcal{O}(\delta F^2,\delta F', \delta G,  \delta w) 
\right]  \,
\end{align}
which coincides to Eq.~\eqref{eq:deltaT-b} after substituting the asymptotic form of $\delta F$, Eq.~\eqref{eq:asymp1_F}.


\bibliographystyle{jhep}
\bibliography{./references}

\end{document}